\documentclass[aip,jcp,superscriptaddress,reprint,floatfix,citeautoscript]{revtex4-1}

\usepackage{geometry}

\usepackage[english]{babel}

\usepackage{amssymb}
\usepackage{amsmath}
\usepackage{amsfonts}
\usepackage{newtxtext}
\usepackage[slantedGreek]{newtxmath}
\DeclareSymbolFont{UPM}{U}{eur}{m}{n}  % otherwise the curly d is italic
\DeclareMathSymbol{\partial}{0}{UPM}{"40}
\usepackage[utf8]{inputenc}
\usepackage{mathrsfs}
\usepackage[mathcal]{euscript}
\usepackage[version=4]{mhchem}
\usepackage{siunitx}
\DeclareSIUnit\Molar{\mole\per\liter}
\DeclareSIUnit\calorie{cal}
\usepackage[stretch=15,shrink=15,step=1]{microtype}
\SetProtrusion{encoding={*},family={*},series={*},size={6,7}}
              {1={ ,550},2={ ,300},3={ ,300},4={ ,300},5={ ,300},
               6={ ,300},7={ ,400},8={ ,300},9={ ,300},0={ ,300}}
\usepackage{setspace}
\usepackage{xcolor}

\newcommand{\nicefrac}[2]{\ensuremath{#1/#2}}

\DeclareUnicodeCharacter{016B}{\~u}
\DeclareUnicodeCharacter{0117}{\.e}
\DeclareUnicodeCharacter{2010}{--}
\DeclareUnicodeCharacter{0160}{\v{S}}
\DeclareUnicodeCharacter{0161}{\v{s}}
\DeclareUnicodeCharacter{212B}{\AA}

\usepackage{checkend}

\usepackage{graphicx}
\usepackage{float}

\usepackage{booktabs}

\usepackage[nolist,nohyperlinks]{acronym}

\usepackage{comment} %\begin{comment} \end{comment}

\usepackage[pdftex, bookmarks, pdffitwindow=false, pdfstartview=FitH, pdfdisplaydoctitle, colorlinks, plainpages=false, pdftitle={Lattice models and Monte Carlo methods for simulating DNA origami self-assembly},pdfauthor={Alexander Cumberworth, Aleks Reinhardt, Daan Frenkel}, pdfpagelabels, hypertexnames, citecolor={blue},linkcolor={red}, urlcolor={violet}, pdflang={en}, hyperfootnotes=false, breaklinks]{hyperref}
\usepackage{cleveref}

\makeatletter
\def\@bibdataout@aip{
 \immediate\write\@bibdataout{
 @CONTROL{
   aip41Control, author="08",editor="1",pages="0",title="0",year="1", eprint=-1
 }}
 \if@filesw
  \immediate\write\@auxout{\string\citation{aip41Control}}
 \fi
}
\makeatother

\usepackage{etoolbox}
\preto\citet{~}

\usepackage{mleftright}

\usepackage{marginnote}

\makeatletter
\providecommand*{\standardstate}{{\ensuremath{\protect\cst@sstate}}}
\newcommand*{\cst@sstate}{\mathpalette\cst@s@state\circ}
\newcommand*{\cst@s@state}[2]{\ooalign{\hfil$#1-$\hfil\cr\hfil$#1#2$\hfil\cr}}
\makeatother

\renewcommand{\vec}[1]{\boldsymbol{#1}}

\newcommand{\kb}{k_{\mathrm{B}}}

\newcommand{\duxy}[2]{\upDelta U(\vec{#1}, \vec{#2})}

\newcommand{\dbxy}[2]{\mathrm{e}^{-\beta \duxy{#1}{#2}}}

\newcommand{\axy}[2]{p_{\mathrm{acc}}(\vec{#1} \mid \vec{#2})}

\newcommand{\ti}{p_i^{\mathrm{trial}}}

\newcommand{\po}{p_{i, j}^{\mathrm{open}}}

\newcommand{\txyin}{p_{\mathrm{trial}}(\vec{x}, N_i + 1 \mid \vec{y}, N_i)}

\newcommand{\txydel}{p_{\mathrm{trial}}(\vec{x}, N_i - 1 \mid \vec{y}, N_i)}

\newcommand{\axyin}{p_{\mathrm{acc}}(\vec{x}, N_i + 1 \mid \vec{y}, N_i)}

\newcommand{\axydel}{p_{\mathrm{acc}}(\vec{x}, N_i - 1 \mid \vec{y}, N_i)}

\newcommand{\fug}{\mathrm{e}^{\beta \mu_i}}

\newcommand{\ifug}{\mathrm{e}^{-\beta \mu_i}}

\begin{document}

    % Metadata
    \title{Lattice models and Monte Carlo methods for simulating DNA origami self-assembly}

    \author{Alexander Cumberworth}
%     \email{amc226@cam.ac.uk}

    \author{Aleks Reinhardt}

    \author{Daan Frenkel}
%     \email{df246@cam.ac.uk}

    \affiliation{Department of Chemistry, University of Cambridge, Lensfield Road, Cambridge, CB2 1EW, United Kingdom}

    % abstract.tex

\begin{abstract}
The optimal design of DNA origami systems that assemble rapidly and robustly is hampered by the lack of a model for self-assembly that is sufficiently detailed yet computationally tractable.
Here, we propose a model for DNA origami that strikes a balance between these two criteria by representing these systems on a lattice at the level of binding domains.
The free energy of hybridization between individual binding domains is estimated with a nearest-neighbour model.
Double helical segments are treated as rigid rods, but we allow flexibility at points where the backbone of one of the strands is interrupted, which provides a reasonably realistic representation of partially and fully assembled states.
Particular attention is paid to the constraints imposed by the double helical twist, as they determine where strand crossovers between adjacent helices can occur.
To improve the efficiency of sampling configuration space, we develop Monte Carlo methods for sampling scaffold conformations in near-assembled states, and we carry out simulations in the grand canonical ensemble, enabling us to avoid considering states with unbound staples.
We demonstrate that our model can quickly sample assembled configurations of a small origami design previously studied with the oxDNA model, as well as a design with staples that span longer segments of the scaffold.
The sampling ability of our method should allow for good statistics to be obtained when studying the assembly pathways, and is suited to investigating in particular the effects of design and assembly conditions on these pathways and their resulting final assembled structures.
\end{abstract}

    \maketitle

    % Acronyms
    % glossary.tex

\begin{acronym}
    \acro{AFM}{atomic force microscopy}
    \acro{bp}{base pair}
    \acro{CB}{configurational bias}
    \acro{CT}{conserved topology}
    \acro{ds}{double stranded}
    \acro{MC}{Monte Carlo}
    \acro{MD}{molecular dynamics}
    \acro{NN}{nearest-neighbour}
    \acro{nt}{nucleotide}
    \acro{REMC}{replica exchange Monte Carlo}
    \acro{RG}{recoil growth}
    \acro{SI}{supplementary information}
    \acro{ssDNA}{single-stranded DNA}
\end{acronym}

    % intro.tex

\section{Introduction}
\label{sec:intro}
In the early 1980s, \citet{seeman1982} demonstrated that DNA, the carrier of genetic information, could also be used as a versatile design material to build novel nano-structures.
Since then, there has been great interest in pushing the limits of the size and intricacy of the structures that can be designed and assembled with DNA.
But it was not until the seminal paper of Rothemund~\cite{rothemund2006}, which introduced the DNA origami method, that the complexity of structural DNA nanotechnology systems really took off.
The key idea behind DNA origami is to employ a long \ac{ssDNA} `scaffold' strand that is subsequently folded into its target structure by hybridising a number of designed, shorter `staple' strands that link selected binding domains on the scaffold strand~\cite{rothemund2006}.
A fully assembled origami structure is largely made up of parallel double helices with strand crossovers between adjacent parallel helices, which provide structural stability.

While early designs were mostly simple planar structures, it is now possible to design and assemble virtually any connected 3D shape, even structures with flexible joints~\cite{douglas2009,*castro2011,*han2013,*benson2015,*castro2015,*zhang2015b}.
This variety of design, and the ability to functionalize the staple strands individually with other molecules, opens up DNA origami to applications including chemical sensors~\cite{ke2008}, nanoreactors~\cite{linko2015b,*liu2013}, electronic devices~\cite{maune2010} and drug delivery vehicles~\cite{linko2015}.
Although the rules for designing a DNA origami system with a particular final structure are well understood, our understanding of the precise assembly thermodynamics and kinetics (i.e.~the order and cooperativity of staple binding) is much more limited.
Yet such understanding is potentially very useful for designing origami structures that fold most efficiently into their target structure.
Given the range of possible applications of DNA origamis, improving the speed and yield of their assembly may have significant practical use.

The factors that determine the kinetics of origami formation are different from those that determine, say, the formation of an ordered crystal.
In fact, \citet{cademartiri2015} differentiate between two fundamentally different types of self-assembly: the \emph{puzzle} mechanism and the \emph{folding} mechanism.
Crystals and many periodic structures typically form via the puzzle mechanism.
The information of how and where a given component must bind under the puzzle mechanism is entirely stored in the interactions between components; they move freely through the solution until the correct partners in the correct orientation are encountered.
To achieve complex and addressable structures~\cite{frenkel2015}, the interactions between components must be highly specific (e.g.~in DNA bricks~\cite{ke2012}).
By contrast, the folding mechanism relies on the fact that some of the components are already covalently bonded to each other, where the covalent bonds are formed by some non-self-assembling process.
As an example, protein folding starts from a structure in which the individual amino acids have been bonded together by a ribosome in a particular sequence that is encoded in the genome.
This pre-forming of more permanent bonds by some other process allows complex structures to be encoded with less specific types of interaction between the individual components (e.g.~non-covalent interactions between amino acids) because of the additional constraints on the system.
With the binding of the staples to specific segments of a scaffold strand and the subsequent folding up of the scaffold strand, DNA origami combines both of these approaches.

The assembly of DNA origami is a cooperative process~\cite{sobczak2012}.
This follows from the fact that the melting and annealing curves, as measured with spectroscopic techniques, are narrower than those of the corresponding isolated binding domains.
The most obvious form of cooperativity is the increase in local concentration of binding domains when they are brought together by a staple that binds nearby domains to form a loop~\cite{dunn2015}.
However, \citet{dannenberg2015} found that coaxial stacking between staples adjacent to each other on the scaffold may also increase cooperativity of the assembly process.
Both forms of cooperativity would be expected to have primarily a local effect and indeed FRET experiments~\cite{wei2013} and simulations~\cite{dannenberg2015} have found that excluding a staple from the reaction mixture only affects the binding of nearby staples in the assembled structure.

An important question about the assembly process is whether it is under kinetic or thermodynamic control, and to what extent this depends on the conditions, assembly protocol and origami design.
Hysteresis between the melting and annealing curves is commonly observed~\cite{dunn2015,dannenberg2015,wei2013,arbona2013,*arbona2012,*arbona2012b}, with annealing occurring at lower temperatures than melting; the effect seems to be stronger in 3D origami structures~\cite{wei2013}.
Increasing the heating/cooling rate has been found to have a more pronounced effect on annealing than on melting ~\cite{dannenberg2015}, which suggests that for some conditions and designs assembly is slow relative to melting and may involve high free-energy barriers, corresponding to a process under kinetic control.
Conversely, \citet{wah2016} examined intermediates of the assembly process with \ac{AFM} of origamis of a similar design and found that the annealing and melting pathways were largely the reverse of each other, and that the calculated melting and annealing temperatures were in agreement within experimental error, suggesting a thermodynamically controlled process.
Using both \ac{AFM} measurements and simulations, \citet{dunn2015} examined an origami design for which multiple fully assembled configurations were equally stable and showed that both thermodynamic and kinetic factors could be manipulated to control the outcome of the assembly reaction.
They were able to shift the assembly yield towards specific assembled configurations by modifying the staple designs to increase the stability of those states.
A similar shift in assembly yield was also able to be achieved with staple modification that instead changed the stabilities of some of the intermediate states in the assembly pathways, leaving the stabilities of the assembled configurations unchanged.

Molecular simulations can be used to gain a better understanding of the factors that influence the thermodynamics and kinetics of assembly and melting of DNA origami structures.
Unfortunately, as in protein-folding simulations, direct atomistic simulations of the assembly process are not feasible for all but the shortest sequences.
It is therefore necessary to use simplified yet realistic models.
In fact, several such models that vary in their level of detail have been introduced.
As we argue below, all existing approaches have their drawbacks, either providing too little information about the pathway, or being too slow for large origamis.

In the lower-resolution models, the standard approach is to model DNA origami self-assembly by extending the \citet{santalucia2004} \ac{NN} model of DNA hybridization to account for the entropic effects of folding the scaffold.
Using this approach, Arbona, Aim\'e and Elezgaray~\cite{arbona2013,*arbona2012,*arbona2012b} modeled the assembly process as a series of equilibrium reactions to calculate the likelihood that a particular staple or individual staple binding domain is bound to its complementary binding domain(s) on the scaffold at a given temperature.
\citet{dannenberg2015} and \citet{dunn2015} instead formulated their model as a continuous-time Markov chain, where the state space is described by the binding states of each staple type in the system.
These approaches allow the assembly process to be simulated in under an hour on current computers.
Nevertheless, the efficiency advantage that these statistical models provide comes at the price of having no explicit geometric representation of the system, and making fairly strong assumptions about the entropic changes that occur during assembly.
Furthermore, these models ignore the possibility that staples may bind (albeit less strongly) to incorrect binding domains.

In the higher resolution approach, \citet{snodin2016} used a coarse-grained model of DNA known as oxDNA~\cite{snodin2015b,sulc2012,ouldridge2011,ouldridge2010} to model the self-assembly of a small DNA origami.
They were able to capture a full assembly event in unbiased simulations of the system with their model, which allowed them to study the process in unprecedented detail.
However, because of the level of detail that oxDNA provides, these simulations of a small origami design with only short loops present in the final structure took several months on a cluster with GPU acceleration.
Moreover, they found it necessary to use staple concentrations in excess of those typically used in experimental assembly conditions.
This is not just a matter of speeding up the kinetics: such high concentrations shift the equilibrium between free and bound staple strands towards the bound states.

Here, we bridge the gap between these two approaches for simulating the assembly of DNA origami by introducing a model that aims to combine the low computational cost associated with the SantaLucia model with the structural information provided by the oxDNA model.

%    \newpage
    % model.tex

\section{Model}
\label{sec:model}

\subsection{Model description}
\label{sec:model-description}

\begin{figure}
    \includegraphics{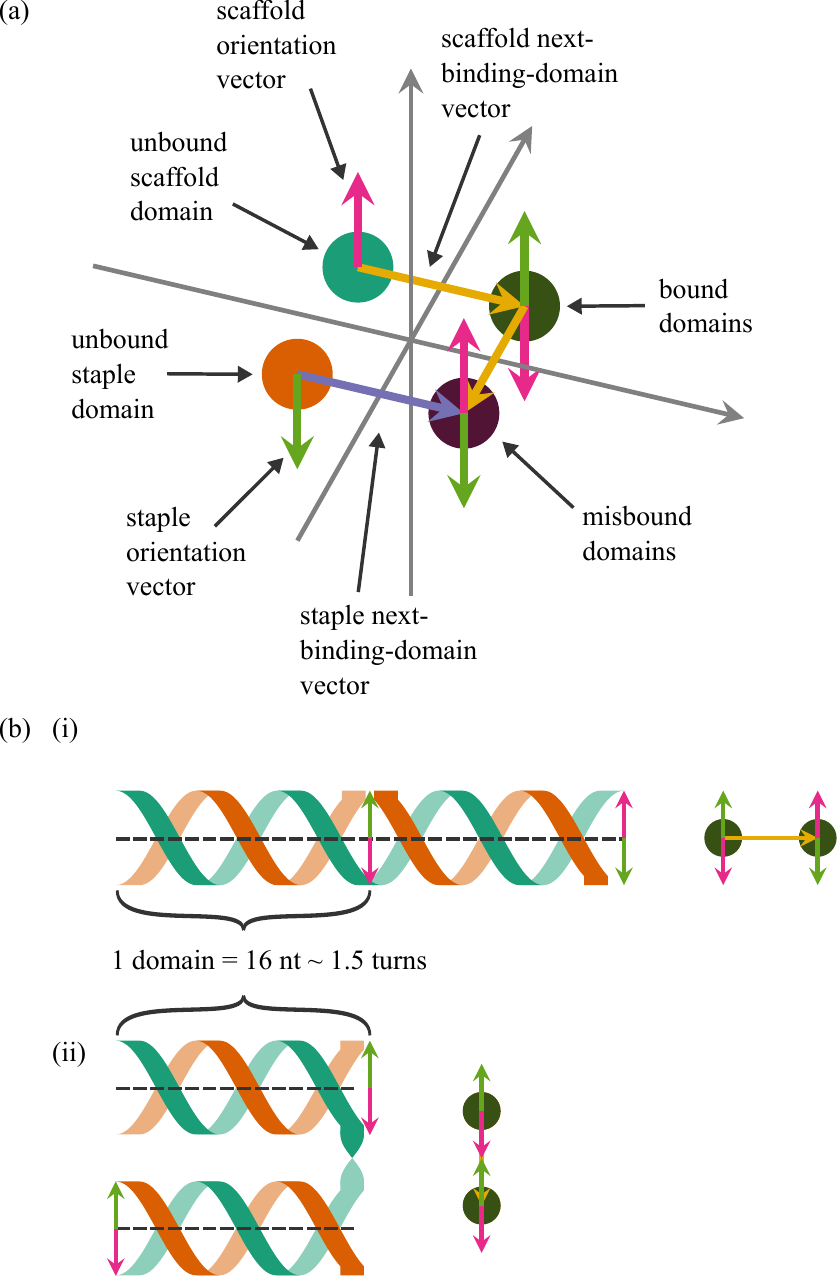}
    \caption{
        \label{fig:overview}
        Schematic illustrations of the basic elements of the model.
        (a) Occupied lattice states.
        There is one scaffold with three binding domains, one staple with two binding domains, and one staple with a single binding domain.
        All vectors shown are unit vectors.
        (b) Two binding domains of a scaffold and a staple in both an idealized cartoon helix representation and the corresponding lattice-model representation.
        The binding domains are 16-\ac{nt} long, which in B-form DNA corresponds to about 1.5 turns of the helix.
        In (i), both binding domains are part of the same helix, while in (ii), they are part of separate helices with a kink between them; this particular kinked configuration is also that of a strand cross-over between two adjacent parallel helices.
    }
\end{figure}

In the context of DNA origami, a `binding domain' is defined as a segment of an individual DNA chain that, in the final assembled state, is fully bound to another, complementary segment of DNA.
In our model, we represent such binding domains as particles on a lattice.
We chose a lattice representation in order to increase computational efficiency by reducing configuration space.
Further, the use of a lattice representation for structural DNA nanotechnology systems has good precedent: a lattice model of DNA bricks was remarkably effective~\cite{reinhardt2014,*jacobs2015,*reinhardt2016}, and unexpectedly yielded near quantitative agreement with experimental measurements of the nucleation kinetics~\cite{sajfutdinow2018}.
For simplicity, in this work we consider origami designs in which the angles between helical axes involve only angles that are multiples of \ang{90} in the final structure, so we chose to work with a simple cubic lattice.
Contiguous binding domains on a given chain are constrained to occupy adjacent lattice sites.
The lattice sites can have an occupancy of zero (unoccupied), one (unbound), or two (bound or misbound), where the number indicates how many domains are present at that site (\cref{fig:overview}(a)).
Bound states are defined to be only those in which the two binding domains occupying the same lattice site have fully complementary sequences; if the sequences are not fully complementary, it is defined as a misbound state.

To account for the fact that DNA strands bond with one another, we compute an energy of interaction for all bound or misbound lattice sites.
This energy of interaction in our model is taken to be the hybridization free energy of the two strands that occupy the same lattice site, accounting not only for the energy of bonding, but also, in a coarse-grained way, for the entropy of hybridizing two molecules.
Consequently, the interaction energies in our model are strongly temperature dependent.
We compute the hybridization free energies associated with bound and misbound states using the SantaLucia \ac{NN} model~\cite{santalucia2004}, which are a function of both temperature and salt concentration.
In the case of partially complementary sequences, the hybridization free energy is approximated by the predicted free energy for the longest contiguous complementary sequence of the pair; this approximation has been shown to work well when simulating DNA bricks~\cite{wayment-steele2017,reinhardt2014,*jacobs2015}.

The primary challenge in designing a model at this level of resolution for the simulation of DNA origami self-assembly is to account for the constraints imposed by the double helical twist on the structure of the system.
There are two aspects to this challenge.
First, we need constraints to restrict where a strand crossover can occur between parallel helices.
This is necessary because the strands of two adjacent parallel helices are only in a position compatible with a strand crossover at certain intervals of base pairs along the helices.
Second, some way of transmitting information on the current phase of the helical twist along adjacent binding domains in the same helix is needed.

By associating an orientation unit vector with each binding domain, it is possible to create a set of rules that meets both requirements.
In arguing for the form of the model, we will refer to diagrams of an idealized double helical structure of DNA, rather than a fully atomistic model, which is sufficient for the level of detail we are targeting in the design of our model.
In a bound or misbound state, we define the orientation vector as the vector which points out orthogonally from the helical axis to the position of the strand at the end of the helix in the current binding domain.
Suppose two particles in our model occupy a given lattice site.
According to the above definition of an orientation vector, if the lattice site is in a bound or misbound state, the orientation vectors of the two binding domains must add up to zero.
As an example, consider a system with two fully complementary pairs of 16-\ac{nt} binding domains, as in \cref{fig:overview}(b)(i).
At the end of the first (leftmost) binding domain, the scaffold strand (teal) is at the bottom of the helix, and so the orientation vector for the binding domain of the scaffold strand (pink arrow) points downwards from the centre line.
By contrast, the staple strand (orange) is at the top of the helix at the same point, and so the orientation vector for the binding domain of the staple strand (green arrow) points upwards from the centre line.
In an unbound state, the direction of the orientation vector is uniformly distributed.

\begin{figure}
    \includegraphics{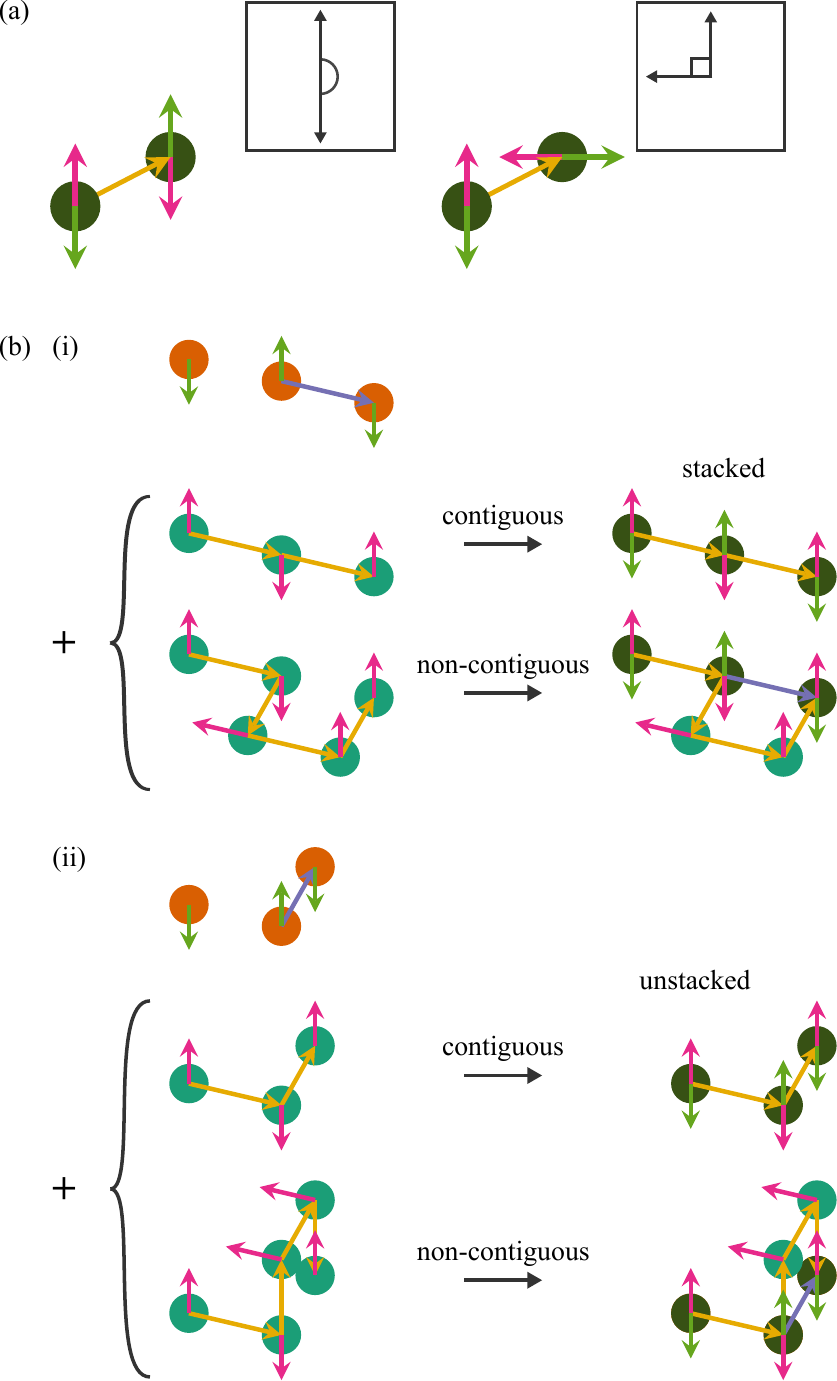}
    \caption{
        \label{fig:stacking}
        Helical stacking in the model.
        (a) Orientation vectors and helical phase.
        The boxes are projections of the scaffold orientation vectors of the two binding domains onto a plane normal to the next-binding-domain vector, with the dihedral angle indicated.
        Left: the orientation vectors are consistent with two stacked 16-\ac{nt} (1.5 turn) binding domains.
        Right: the orientation vectors are consistent with two stacked 8-\ac{nt} (0.75 turn) binding domains.
        (b) Three bound domain pairs in which each pairing of contiguous domains along the strand is in a stacked configuration (all binding domains are 16~\ac{nt} in length).
        The staples and scaffold are shown separately for clarity, and a scaffold configuration is given for both the case in which there is and the case in which there is not one contiguous segment of a chain forming the three bound domain pairs.
        (i) Stacked triplet configurations.
        (ii) Unstacked triplet configurations.
    }
\end{figure}

The orientation vector thus clearly contains information about the current phase of the twist at the end of the helix in the binding domain.
In order for the model to be consistent with helical geometry, which here is assumed to correspond to B-DNA, when two adjacent binding domains are in the same helix, the dihedral angle between the planes defined by the orientation vectors and the vector connecting the two domains must be determined by the number of turns of the helix between them (\cref{fig:stacking}(a)).
We shall refer to the unit vector that connects two binding domains as the `next-binding-domain vector', since for a given binding domain, it points to the next binding domain along the chain.
In the case of a scaffold chain, the positive direction is defined as 5' to 3', while in the case of a staple chain, it is defined as 3' to 5'.

The dihedral angle that we expect depends on the length of the binding domains.
For example, in the case we considered above (\cref{fig:overview}(b)), each binding domain corresponds to 1.5 turns of the helix.
If two adjacent bound domains are in the same helix, we therefore expect a dihedral angle of \ang{180}; the orientation vectors of the scaffold (pink) and staple (green) strands in \cref{fig:overview}(b)(i) must therefore alternate in sign when they are part of the same helix.
We also note that if two contiguous domains are in bound states and part of the same helix, then there is an additional `stacking' interaction (see \cref{sec:model-parameters} for further discussion of this interaction) that we have not yet accounted for when calculating the \ac{NN} model hybridization free energy for the two domains separately.

The length of the binding domains is typically well below the persistence length of double stranded DNA, so we need to ensure that helices are rigid in this model.
Because there is no explicit helical axis vector in the model, a single binding domain pair will only implicitly define the helical axis to lie within a plane.
The helical axis is not resolved until an adjacent binding domain enters a bound state in the same helix.
Explicit checks of helical rigidity are thus only necessary once we consider triplets of bound domain pairs.
Consider two pairs of adjacent lattice sites with one site in common, where all three sites are in bound states.
If for each pair, the two sites are occupied by two contiguous binding domains of a chain, and the two contiguous bound pairs satisfy the same helix geometry, then in order for all three bound domain pairs to be in the same helix, the lattice sites must all lie on a common line (\cref{fig:stacking}(b)(i)).
These two pairs of two contiguous binding domains could either be a single contiguous segment of three domains, or they could be two separate segments that are bound to each other at the middle lattice site.

The above considerations do not apply if one of the domains is misbound, because regions of ssDNA, which has a much lower persistence length, are then present between the helical sections.
Points in a helix at which there is a break in the backbone of one of the strands, which we will refer to as breakpoints, are also more flexible, and are able to become unstacked to form kinks.
We account for the flexibility at breakpoints in our model by allowing such kinks to form.
In the model, if the orientation vectors of a pair of contiguous bound domains do not have a configuration prescribed by the helical geometry, they are considered to have a kink and are treated as two separate helices with no stacking interaction (for an example see \cref{fig:overview}(b)(ii), which also happens to be a crossover configuration).
If a triplet of bound domains of the type described above does not meet the conditions necessary for it to be a single helix, then the configuration is likewise considered to have a kink between one of the contiguous binding domain pairs (\cref{fig:stacking}(b)(ii)).
These configurations will have one less stacking interaction than configurations in which all three bound domain pairs are in the same helix.

Consider a pair of contiguous bound domains with a kink between them.
In reality, the kink will not allow for all possible relative orientations of the two binding domains.
By considering transformations to the two helical domains and making simple steric arguments, we can introduce two further rules (see \cref{sec:appendix-kinks} for detailed arguments).
First, if the first binding domain's next-binding-domain vector is perpendicular to its orientation vector, the second binding domain's orientation vector must also be perpendicular to the next-binding-domain vector of the first.
Second, if the first binding domain's next-binding-domain vector is parallel to the first binding domain's orientation vector but not parallel to the second binding domain's orientation vector, or if the first binding domain's next-binding-domain vector is antiparallel to its orientation vector, then the configuration is disallowed.
If both of the bound domains with a kink between them have a defined helical axis because of an adjacent bound domain (on either the chain with the kink or on one of the chains bound to the domains forming the kink), a third rule is applied: if the two helical axes are parallel or antiparallel to each other and orthogonal to the next-binding-domain vector between the binding domains forming the kink, then there is an additional kink present amongst these four bound domains and thus one less stacking interaction.
These rules also happen to ensure that the first aspect of the challenge to reproduce the constraints of the double helical twist is met, namely that crossovers only occur at certain intervals between parallel helices.

\begin{figure*}
    \includegraphics{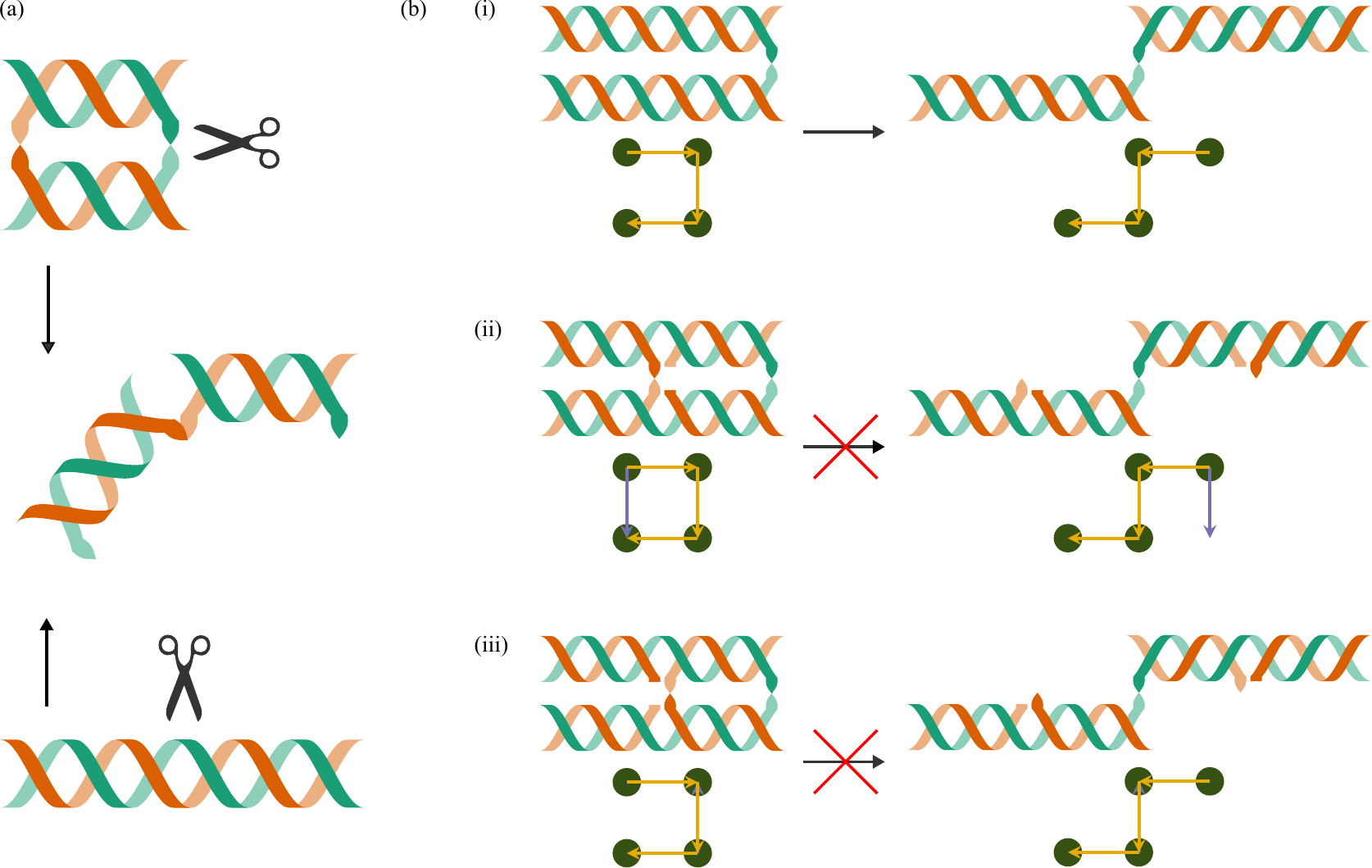}
    \caption{
        \label{fig:crossovers}
        Strand crossovers between helices involving 16-\ac{nt} long binding domains.
        For clarity, the orientation vectors have been omitted.
        (a) Doubly contiguous binding domains in bound states.
        (b) Crossovers between two adjacent parallel helices with a four-binding-domain scaffold.
        Red crosses over transformation arrows imply that the transformation cannot be achieved because a covalent bond would need to be broken.
        (i) Helices with a single crossover.
        (ii) Helices with two crossovers on separate binding domains.
        (iii) Helices with crossovers on the same binding domain.}
\end{figure*}

Consider two adjacent lattice sites in bound states, where at least one pair of binding domains are contiguous.
If the other pair of binding domains are not contiguous and in the same helix, their orientation vectors will still satisfy the prescribed helical angle because of the requirement of their orientation vectors to be opposing those of the strand that has two contiguous binding domains in that helix.
However, the case in which both pairs of binding domains are contiguous requires further consideration.
In reality, if the combined sequence of the two binding domains on one chain is together the reverse complement of the combined sequence of the two binding domains on the other chain, then the only way for all binding domains to be bound to each other is if there is only one helix.
If instead the binding domains on one chain must be swapped to make the whole two-binding-domain sequence the reverse complement of the other whole two-binding-domain sequence, then the only way for all binding domains to be bound to each other is if there are two parallel helices with both strands crossing over.
As a concrete illustration, one of the chains would have to be cut and glued to its other end to transition between these two configurations (\cref{fig:crossovers}(a)).
Thus, the model constrains pairs of contiguous complementary binding domains bound to each other to be in the same helix if they are the full reverse complements of each, and to be crossing over if not.

In reality, when there is more than one crossover between two helices, the helices become much more restricted in the configurations they are able to take relative to each other (compare \cref{fig:crossovers}(b)(i) to (ii)).
In particular, they will be forced to be roughly parallel.
This is naturally captured by the model when there are crossovers between more than one set of binding domain pairs on two separate helices, as seen in \cref{fig:crossovers}(b)(ii).
However, when a binding domain pair is involved in a double crossover as discussed above, which can occur in binding domains with a length of 16~\ac{nt}, this will not be captured by the model as currently defined (\cref{fig:crossovers}(b)(iii)).

To impose the parallel helices constraint on the system in the case of a double crossover between two binding-domain pairs, one must consider only cases where there are two additional contiguous binding domains on one of the chains involved, one before and one after the pair forming the double crossover, where both have orientation vectors that allow them to be in the same helix as the relevant binding domain they are contiguous with.
Then, in order for the pairings before and after the crossover to both be in the same helix, the next-binding-domain vector of the first binding domain and the next-binding-domain vector of the third binding domain must be antiparallel, or equivalently, the first and last binding domains must occupy adjacent lattice sites.
If this is not the case, the configuration has one more kink, and thus one less stacking interaction, than configurations where these criteria are met.
Further, configurations where the first and last binding domains are on opposing sides of the junction, or more precisely, where the next-binding-domain vectors are parallel, are sterically prohibited, and so are made disallowed configurations in our model.

\subsection{Initial parameter selection}
\label{sec:model-parameters}

There are three main parameters related to the assembly conditions that must be chosen in order to run a simulation: the staple concentration, the salt concentration and the temperature.
Here, we set all staples to have the same amount concentration $C$.
While staple concentrations used vary from study to study, typical values are around \SI{100}{\nano\Molar}, which is the value we use here.
When mapping this concentration to the chemical potential for simulations in the grand canonical ensemble, in order to be consistent with hybridization experiments, we assume that we can write the staple-strand chemical potential to within a constant as $\mu_i = \kb T \ln(C/C^\standardstate)$, where $C^\standardstate=\SI{1}{\Molar}$.
Assembly is typically carried out in solutions with significant amounts of dissolved salts, so we set the monovalent cation concentration to be \SI{0.5}{\Molar}, which is the same concentration as that used by \citet{snodin2016} in their oxDNA simulations.
As we are running \ac{REMC} simulations (\cref{sec:methods}), we do not have to select a single temperature; the range is selected to encompass the region of the relevant order parameter curves with the steepest gradient, approximately centred on the melting temperature.

The physical origin of the stacking interactions is the attractive $\uppi$--$\uppi$ interactions between the aromatic rings of the bases of adjacent nucleotides.
Since stacked states have a constrained geometry, they are entropically disfavoured.
However, because some of the entropic component of the stacking free energy is accounted for by constraining the orientation vectors in bound states within our model, we assume the stacking interaction parameter is entirely energetic and treat the stacking interaction as a single temperature-independent tunable parameter.
To select its value, we ran short serial simulations of a two-binding-domain scaffold with two single-binding-domain staples at a temperature below the melting temperature and calculated the free-energy difference between the stacked and unstacked states for a range of stacking energies (\cref{fig:stacking-enthalpy}).
We selected a value of $-1000\,k_\text{B}\si{\kelvin}$, which gave a free-energy difference that roughly matched experimentally measured values~\cite{protozanova2004,*yakovchuk2006} of $\sim$\SI{-4}{\kilo\joule\per\mole}.

In the next section, we describe the Monte Carlo scheme that we have used to simulate origami self-assembly.
It turns out that, in order to achieve good efficiency, we need to use a combination of various more advanced sampling schemes that we describe below.
However, readers who are primarily interested in the results of our simulations may prefer to go directly to \cref{sec:results}.

%    \newpage
    % methods.tex

\section{Monte Carlo methods}
\label{sec:methods}

The details of the behaviour of staple strands in the assembly process are of interest solely when they are (mis)bound to a scaffold strand.
Only the availability of the staples for binding to scaffold strands is relevant; this availability is determined by the initial staple concentrations, the binding of staples to scaffolds, and the binding of staples to other staples.
Because in a typical assembly protocol, the staples are present in excess of the required stoichiometry, it can be assumed to a first approximation that the free staple concentrations are constant over the course of the assembly process.
Further, at the temperatures relevant to assembly, staple--staple binding should not be a significant factor because the staples are not designed to bind to each other.
The sampling of states with free staples can be avoided by running the simulations in the grand canonical ensemble, in which we fix the chemical potential of the staples rather than their number.
While staple--staple binding is not favourable overall, because of the local increase in concentration of staples at the scaffold, we do allow staple--staple binding to occur.
Thus in states with no free staples, the staples can either be (mis)bound directly to the scaffold or (mis)bound indirectly via binding to a staple already (mis)bound to the system.

In order to increase sampling efficiency, we use \ac{REMC}~\cite{marinari1992}.
This advanced sampling method involves running multiple replicas of the simulation that differ only with respect to a few of the simulation parameters.
While we are primarily interested in running the simulations across a range of temperatures, here we must also use a range of Hamiltonians because of the temperature dependence of the \ac{NN} model hybridization free energies.
The simplest form of \ac{REMC} involves an exchange attempt at a random step interval of the configurations of a random pair of replicas, provided that the replicas are adjacent with respect to the variable that distinguishes different replicas.
To improve parallelization, the variant of \ac{REMC} used here attempts an exchange at a set step interval, and alternates between attempting an exchange between all even pairs and all odd pairs of replicas, where pairs are numbered with the index of the first replica in the pair along the exchange variable.
While this variant of \ac{REMC} no longer obeys detailed balance, it can be shown to obey total balance~\cite{manousiouthakis1999}.

The strong and specific interactions of the model and the need to sample states with different numbers of (mis)bound staples makes efficient sampling challenging.
To meet the challenge, we developed several \ac{MC} move types, which we outline in the following subsections.
While they are described as applied to the current DNA origami model, these move types are applicable in general to polymer lattice models.

\subsection{Monte Carlo move types}
\label{sec:methods-movetypes}

All move types described below are dynamic (i.e.~the move types use the current configuration to generate a trial configuration, thus producing configurations that are correlated but not necessarily in a way that mimics the dynamics of real origami assembly) and obey detailed balance.
Apart from those move types that only modify orientation vectors, they involve sequential growth of a set of binding domains.
If the move involves regrowth of binding domains already present in the system volume, they are first unassigned (i.e.~the lattice sites become unoccupied) before regrowth begins.
The order in which they are grown depends on the specifics of the move type, but in the simplest case, the set of binding domains to be grown forms a contiguous segment of a single strand, and the binding domains are grown according to their order in the strand.
More complex move types may involve multiple segments from the same strand, or segments from multiple strands, where growth of binding domains from a given segment in the stack may be interrupted by growth of binding domains from other segments.
Growth requires selection of a position on a neighbouring lattice site and an orientation vector.
The position can be decomposed into the sum of the position of the lattice site being grown from and a unit vector, which we refer to as the position difference vector.

Growth is done with three main variants: symmetric, \ac{CB} and \ac{RG}.
In the symmetric variants, the position difference vectors and orientation vectors are chosen with uniform probability from the set of all unit vectors.
As generation of configurations is symmetric, the trial generation probabilities of binding domain growth for the forward and reverse moves will cancel.
Thus the move is accepted according to the classic canonical Metropolis acceptance probability,
\begin{equation}
    \label{eq:met}
    \axy{x}{y} = \min[1,\ \dbxy{x}{y}],
\end{equation}
where $\duxy{x}{y}$ is the change in energy from the old configuration $\vec{y}$ to the new configuration $\vec{x}$, and $\beta$ is the inverse thermodynamic temperature.

In the \ac{CB} variants~\cite{siepmann1992}, the selection of a new configuration for each binding domain is biased by the associated energy change, such that the trial generation probability of each binding domain is
\begin{equation}
    \label{eq:cbtrial}
    \ti = \frac{\mathrm{e}^{-\beta \upDelta \varepsilon_{i, j}}}{\sum_{j'}^{k} \mathrm{e}^{-\beta \upDelta \varepsilon_{i, j'}}},
\end{equation}
where the sum is over the number of possible configurations $k$, which here is the number of neighbouring lattice sites times the number of possible orientation vectors (thus $k=36$), and $\upDelta \varepsilon_{i, j}$ is the energy change of setting binding domain $i$ to have configuration $j$ after having grown out all previous binding domains.
As the trial generation probability is no longer symmetric, the acceptance probability will have additional terms that account for this.
Rearranging and grouping these terms gives
\begin{equation}
    \label{eq:cbaccept}
    \axy{x}{y} = \min \mleft[1,\ \frac{W_\mathrm{new}}{W_\mathrm{old}} \mright],
\end{equation}
where the Rosenbluth weight $W$ is defined in terms of the Rosenbluth weights of each of the $n$ binding domains grown, $w_i$, as
\begin{equation}
    \label{eq:rosenbluth}
    W = \prod_i^n w_i = \prod_i^n \left( \sum_{j = 1}^{k} \mathrm{e}^{-\beta \upDelta \varepsilon_{i, j}} \right).
\end{equation}
$W_\mathrm{old}$ is calculated by growing the old configuration and calculating $w_i$ at each step.

Finally, in the \ac{RG} variants~\cite{consta1999a,consta1999b}, if growth becomes stuck, the binding domains that were previously set can be unassigned, allowing them to be regrown in a different configuration.
The growth of each binding domain involves selecting a configuration with uniform probability and choosing whether to consider the configuration open or not according to some probability distribution.
If it is chosen to be open, the configuration is selected for use and growth of the next binding domain proceeds.
The probability of a configuration being open can be defined as needed; one possibility is
\begin{equation}
    \label{eq:popen}
    \po = \min[1,\ \mathrm{e}^{-\beta \upDelta \varepsilon_{i, j}}].
\end{equation}
If the configuration is chosen to be closed, another is proposed, up to a total of $k_{\mathrm{max}}$ configurations.
If no open configurations result, growth recoils to the previously set binding domain and testing continues for choosing open configurations where it left off.
Recoiling can occur $l_\mathrm{max}$ times, or until all binding domains being grown have been unassigned, which if reached will result in the move being rejected.

To calculate the acceptance probability, the number of available configurations $m_{i, j}$ at each binding domain $i$ in the selected configuration $j$ in both the new and old configuration must be determined.
A binding domain configuration is considered available if there is at least one open configuration for the next $l_{\mathrm{max}}$ binding domains to be grown, or for all the remaining binding domains to be grown if this is less than $l_{\mathrm{max}}$.
For each binding domain in the grown segment, one available configuration is already known; checking for available configurations continues until a total of $k_{\mathrm{max}}$ configurations have been tested.
The \ac{RG} weights are defined as
\begin{equation}
    \label{eq:rgweight}
    W = \prod_i^n w_i = \prod_i^n \left( \frac{m_{i, j}}{\po} \right).
\end{equation}
Then, the move is accepted with
\begin{equation}
    \label{eq:rgaccept}
    \axy{x}{y} = \min \mleft[1,\ \dbxy{x}{y} \frac{W_\mathrm{new}}{W_\mathrm{old}} \mright].
\end{equation}
A super-detailed balance argument is used by \citet{consta1999a} to show that detailed balance is obeyed with this acceptance probability.

\subsubsection{Orientation vector moves}
\label{sec:methods-move types-orientation}

The first step in an orientation vector move consists of selecting a binding domain in the system and generating a new orientation vector, both with uniform probability.
If the binding domain is in an unbound state, the change in energy upon a change in the orientation vector is zero, so the acceptance probability will be unity.
If the binding domain is in a (mis)bound state, the orientation vector of the partner binding domain will also be modified in the trial configuration to be the additive inverse of the proposed orientation vector of the selected binding domain.
This is then accepted according to \cref{eq:met}.

\subsubsection{Staple regrowth moves}
\label{sec:methods-move types-stapleregrowth}

Staple regrowth consists of first either selecting a staple in the system with uniform probability or rejecting the move if no staples are present.
If this staple is a connecting staple, that is, a staple that if removed would leave a network of staples that has no connection to the scaffold, the move is rejected immediately.
Otherwise, one of the binding domains on the selected staple that is in a (mis)bound state is selected to act as a point from which the remainder of the staple will be grown out from.
Then the staple is grown out in both directions with the \ac{CB} method, although in principle any of the growth schemes discussed in \cref{sec:methods-movetypes} may be used.

This scheme introduces an asymmetry into the generation of trial configurations.
The probability of generating a trial configuration involves a factor of $\nicefrac{1}{b_j}$, where $b_j$ is the number of binding domains on staple $j$ (mis)bound to other chains.
This comes from the selection of a binding domain to grow out from.
As $b_j$ can change between the current and trial configuration, there is an additional factor of $\nicefrac{b_j^{\mathrm{old}}}{b_j^{\mathrm{new\vphantom{old}}}}$ in the acceptance probability.
In the case of \ac{CB}, this gives
\begin{equation}
    \label{eq:cbacceptovercount}
    \axy{x}{y} = \min \left[1,\ \frac{W^{\vphantom{\mathrm{old}}}_\mathrm{new\vphantom{j}}}{W^{\vphantom{\mathrm{old}}}_\mathrm{old\vphantom{j}}} \times \frac{b_j^{\mathrm{old}}}{b_{j\vphantom{\mathrm{old}}}^{\mathrm{new\vphantom{l}}}} \right].
\end{equation}

\subsubsection{Staple exchange moves}
\label{sec:methods-move types-exchange}

Staple exchange starts with a uniform random selection of either a staple insertion or staple deletion move.
Then, in either case, a staple type is selected with uniform probability.
While \ac{CB} and \ac{RG} variants could be used, we only use the symmetric scheme for binding domain growth in the insertion move type.
Clearly though, the trial configuration generation probabilities of the forward and reverse moves will not cancel, as there are many ways to insert and grow a given staple, but just one way to remove it.

In the case of an insertion move, a lattice site in the system volume, $V_{\mathrm{sys}}$, which is defined as all the lattice sites occupied by at least one binding domain in the system, is selected with uniform probability to insert the first binding domain of the staple into, leading to a factor of $\nicefrac{1}{V_{\mathrm{sys}}}$ in the trial configuration generation probability.
A binding domain on the staple being inserted is then selected with uniform probability to grow from, leading to an additional factor of $\nicefrac{1}{n_j}$ in the trial probability, where $n_j$ is the length of staple $j$.
Because when inserting into a lattice site with an unbound domain there is only one orientation that will not lead to rejection, this orientation can be selected immediately.
The staple is then grown out from this binding domain, which gives a further factor of $\nicefrac{1}{6^{2(n_j - 1)}}$ to the trial probability.
However, states which involve binding of multiple binding domains to other chains will be over-counted with the current scheme, as there are $b_j$ ways to grow these configurations.
This can be corrected by multiplying the trial probability by a factor of $b_j$.
Altogether, the trial probability of insertion for staple type $i$ is
\begin{equation}
    \label{eq:trial_insertion}
    \txyin = \frac{b_j}{6^{2(n_j - 1)} n_j V_{\mathrm{sys}}},
\end{equation}
where $N_i$ is the number of staples of type $i$.
For a deletion move, a staple of the selected type in the system is selected with uniform probability and removed if it is not a connector (see \cref{sec:methods-move types-stapleregrowth}), which gives a trial probability of 
\begin{equation}
    \label{eq:trial_deletion}
    \txydel = \frac{1}{N_i}.
\end{equation}
Because the number of staples is changing, the probability of being in a particular state is given by the grand-canonical probability distribution.
Using the above trial probabilities, the acceptance probability for insertion of staple type $i$ is
\begin{align}
      \axyin & \\= & {} \min \mleft[ 1, \frac{6^{2(n_j - 1)} n V_{\mathrm{sys}}}{b_j (N_i + 1)} \fug \dbxy{x}{y} \mright],\notag
\end{align}
while for deletion it is
\begin{align}
      \label{eq:deletion}
    \axydel  & \\ = & {} \min \mleft[ 1, \frac{b_j N_i}{6^{2(n_j - 1)} n V_{\mathrm{sys}}} \ifug \dbxy{x}{y} \mright],\notag
\end{align}
where $\mu_i$ is the chemical potential of staple type $i$.

\subsubsection{Scaffold regrowth moves}
\label{sec:methods-move types-scaffoldregrowth}

A seemingly straightforward way to sample scaffold conformational states would be to select a segment of the scaffold and regrow these binding domains and any (mis)bound staples.
However, even with advanced polymer growth schemes like \ac{CB} and \ac{RG}, if the scaffold segment to be regrown is in a near assembled state, the proposed configurations will rarely have a similar number of (mis)bound domains, and will thus be of a substantially less favourable energy.
Hence such moves will thus almost always be rejected.
To address this, we have chosen to keep the sampling of binding states and scaffold conformational states separate by developing variants of \ac{CB} and \ac{RG} that allow the binding state of the system to be left unchanged when regrowing parts of the system, leaving sampling of binding states to the staple exchange and regrowth moves.
Such a separation also simplifies the calculation of the trial generation probabilities by removing the asymmetries involved with changing binding states.
If the system is considered as a network where (mis)bound domain pairs act as nodes, these moves can be thought of as holding the network topology constant, and are thus referred to as \ac{CT} moves.

Fixed-end \ac{CB} is a scheme that allows polymers to be grown to a predetermined endpoint~\cite{dijsktra1994}.
This works by introducing a further bias into the selection of configurations for the growth of each polymer unit.
The bias is the number of ideal random walks from the trial polymer unit's position to the endpoint position, given the number of polymer units remaining to be grown.
Importantly, if a configuration for the polymer unit currently being grown has no ideal random walks available to reach the endpoint, the configuration will have zero probability of being proposed.

Here, we use a similar idea but extend it to allow multiple endpoints per segment, and growth of multiple segments on possibly multiple chains.
When a move involves growing multiple segments, each can have its own set of endpoint constraints.
Once a particular endpoint is reached, the associated endpoint constraint has been satisfied, and so becomes inactive.
If a binding domain that is to serve as an endpoint must also be grown, the endpoint constraint is inactive until the associated binding domain's configuration has been set.

Because of these cases of endpoint positions being set during growth, the number of ideal random walks can no longer be directly used in the bias.
This is because the initial number of ideal random walks for such endpoints could differ between the old and new configurations, and would thus not cancel when taking the ratio of the Rosenbluth weights for the old and new configuration, as it does in the original method.
Instead, we use an indicator function, $\chi_\text{I}(\upDelta \vec{r}_{l, j}, n_{l, i})$, that is unity if walks remain and zero otherwise, where $\upDelta \vec{r}_{l, j}$ is the difference vector between the trial position of configuration $j$ and the position of endpoint $l$, and $n_{l, i}$ is the number of binding domains remaining to be grown between binding domain $i$ and the binding domain of endpoint $l$.
Whether walks remain or not can be determined by checking if the sum of the absolute values of the components of the position difference vector is greater than or equal to the number of binding domains remaining to be grown.

While the endpoint constraints ensure that the system will still have the (mis)bound pairs it began with (with the exception of same-chain misbinding; see following discussion), they do not prevent new pairings from forming.
To prevent new pairings, another indicator function, $\chi_\text{B}(s)$, of lattice site $s$ can be used.
This function is unity if the lattice site is unoccupied, the position of an endpoint of an active endpoint constraint on the segment being grown, or occupied by another binding domain of the chain currently being grown, and zero otherwise.
We allow misbinding between binding domains on the same chain because the staple exchange and regrowth moves will not allow sampling of states involving scaffold binding domains misbinding with themselves.
Because we allow these misbound pairings to form, we must also allow them to unform, and so they are not used by endpoint constraints.
Because misbinding interactions are relatively weak, decreasing the number of misbound pairs will not typically lead to large unfavourable energy changes.
Further, because they are by definition not as specific as fully bound pairs, there are many ways to propose moves that have the same number of misbound pairs.
Finally, we note that changing the number of intra-chain misbound pairs will not introduce asymmetry into the trial probability, as there is no selection of a domain to grow out from involved.

The trial probability of selecting a configuration for binding domain $i$ is now
\begin{equation}
    \label{eq:ctcbacc}
    \ti = \frac{\mathrm{e}^{-\beta \upDelta \varepsilon_{i, j}} \chi_\text{B}(s_j) \prod_l \chi_\text{I}(\upDelta \vec{r}_{l, j}, n_{l, i})}{\sum_{j'}^{k} \mathrm{e}^{-\beta \upDelta \varepsilon_{i, j'}} \chi_\text{B}(s_{j'}) \prod_{l'} \chi_\text{I}(\upDelta \vec{r}_{l', j'}, n_{l', i'})},
\end{equation}
and the Rosenbluth weight is
\begin{equation}
    \label{eq:ctcbrosenbluth}
    w_i = \sum_{j = 1}^{k} \mathrm{e}^{-\beta \upDelta \varepsilon_{i, j}} \chi_\text{B}(s_j) \prod_l \chi_\text{I}(\upDelta \vec{r}_{l, j}, n_{l, i}).
\end{equation}
A similar modification can be made to the \ac{RG} scheme for growing binding domains to construct a \ac{CT} variant.
The modification is made to the probability of a configuration being open,
\begin{equation}
    \label{eq:ctrgpopen}
    \po = \min[1,\ \mathrm{e}^{-\beta \upDelta \varepsilon_{i, j}} \chi_\text{B}(s_j) \prod_l \chi_\text{I}(\upDelta \vec{r}_{l, j}, n_{l, i})].
\end{equation}

The \ac{CT} move types, whether \ac{CT}\ac{CB} or \ac{CT}\ac{RG}, begin with the selection of a segment or segments of the scaffold to regrow.
Then, of the set of staples that are involved in the network of staples (mis)bound to the selected scaffold segment(s), it must be determined which will be regrown and which will act as endpoints.
Here, if a set of staples is involved in a network that includes scaffold binding domains external to the selected segment(s), the staples will remain in their current configuration, with those that are (mis)bound to the selected scaffold segment acting as endpoints for its regrowth.
If a staple is not involved in such a network, it is regrown with the scaffold, with endpoints for the required endpoint constraints being determined during regrowth.

If the scaffold segment was regrown fully before regrowing any of the staples to be regrown, it would result in binding domains on the scaffold being used in endpoint constraints for the staples to be regrown.
However, this would be less effective than regrowing the staples first such that the endpoints were instead on the staples and used by endpoint constraints applied to regrowth of the scaffold, as typically the staples are only two or three binding domains and so often have no way of reaching an endpoint on a scaffold binding domain.
Thus, as the scaffold is regrown, if there is a fully unset staple to be (mis)bound to the binding domain that has just been set, regrowth of the current chain will be put on hold to regrow this staple.
This may also happen while regrowing a staple, in a recursive manner.
If a binding domain on the chain being regrown is to be (mis)bound to an unassigned binding domain on a chain already in the process of being regrown, an endpoint constraint is set up for this other chain (typically the scaffold).

There are many ways to select the set of scaffold segments for regrowth.
We use two variants: single- and multiple-segment selection.
In the single segment, or contiguous scaffold regrowth, variant, the segment is selected such that the distribution of lengths is uniform, where the range of possible lengths is a parameter of the move type.
To create a segment, a uniformly random scaffold binding domain is selected from the set of all scaffold binding domains to act as the seed binding domain from which to create the segment.
A direction with which to add binding domains to the segment is then selected with uniform probability.
Binding domains are added until either the selected segment length or the end of the chain is reached.
If the end of the chain is reached, segments will begin to be added from the other side of the seed binding domain.

In the multiple segment, or non-contiguous scaffold regrowth, variant, the intention is to allow the selection of binding domains for regrowth to be able to jump at points where two scaffold binding domains are adjacent due to a linking staple.
For each move, a maximum possible total number of binding domains to regrow across all segments is chosen with uniform probability, where the range from which the selection is made is a parameter of the move type.
For each individual segment that is created, a maximum segment length is selected with uniform probability, where the range from which the length is selected is another parameter of the move type.
The addition of binding domains to a given segment proceeds until either the maximum segment length is reached, the maximum regrowth length is reached, the end of the chain is reached, or the binding domain following the binding domain being considered for addition to the segment is already part of another segment to regrow.
This last condition is to simplify the construction of endpoint constraints.

Segment creation begins with the selection of a seed scaffold binding domain and direction from which to add binding domains to the segment in the same manner as with the contiguous scaffold regrowth variant.
As binding domains are added to the segment (and all subsequent segments created), if a binding domain is bound to a staple that is bound to another binding domain of the scaffold that is neither already in a segment to regrow nor contiguous with a scaffold binding domain that is in a segment to regrow, it is added to a queue of potential segment seed binding domains.
Once addition of binding domains to the segment has been terminated, if the maximum possible total number of binding domains to regrow has not been reached, a new segment is created with a binding domain from the front of the aforementioned queue.
The direction from which to proceed is selected as with the first segment.
Once addition of binding domains to this segment has been terminated, and if the maximum possible total number of binding domains to regrow has not been reached, a segment beginning from the binding domain in the opposite direction of the previous segment seed will be used as the seed for a new segment, if it exists and if the following domain is not already part of another segment to regrow.
Once addition of binding domains to this segment has been terminated, the steps after initial segment creation are repeated until either the queue is empty or the maximum possible total number of binding domains to regrow is reached.
The segments are regrown in the order in which they were created.
Because the probability of selecting a particular set segments of scaffold binding domains does not depend on the conformation or on whether or not binding domains are misbound to other binding domains on the same chain, the move type obeys detailed balance.

In \cref{sec:appendix-validation}, we provide the details of our numerical validation, and in \cref{sec:appendix-optimization}, we discuss the optimization of the move sets.

%    \newpage
    % results.tex

\section{Results and discussion}
\label{sec:results}

\begin{figure}
    \centering
    \includegraphics{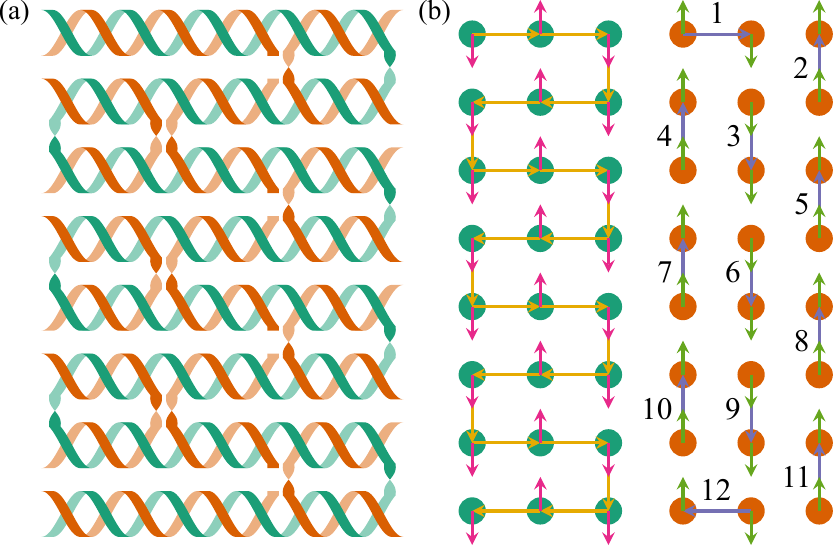}
    \caption{
        \label{fig:snodin-diagram}
        Schematic representations of the 24-binding-domain scaffold system.
        (a) Helical cartoon representation of the system in an assembled, planar configuration.
        (b) Representation of the system with the proposed model.
        The scaffold (left) and staples (right, numbered) are shown in the assembled, planar configuration, but for clarity have been drawn separately.
    }
\end{figure}

\begin{figure}
    \includegraphics{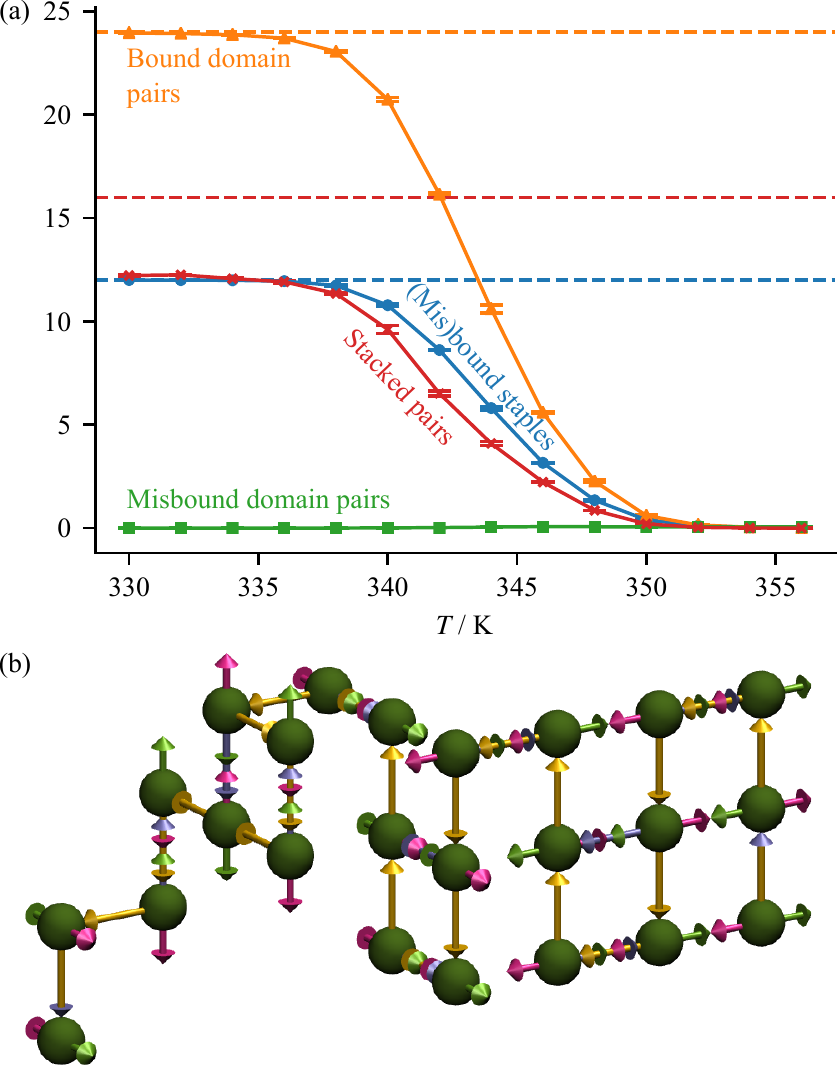}
    \caption{
        \label{fig:snodin-melting}
        Mean order parameters as a function of system temperature and a typical assembled configuration of a 24-binding-domain scaffold system.
        Simulations were run with a staple concentration of \SI{100}{\nano\Molar}, a monovalent cation concentration of \SI{0.5}{\Molar}, and a stacking energy of $-1000\,k_\text{B}\si{\kelvin}$.
        (a) Mean order parameters plotted against temperature.
        The dashed lines correspond to the expected order parameter values in the assembled or fully stacked assembled configurations.
        The error bars represent the standard error in the means across three independent simulations.
        (b) An assembled configuration at \SI{330}{\kelvin}.
    }
\end{figure}

To test the efficacy of our model, we ran simulations of a 24-binding-domain scaffold system previously studied with the oxDNA model~\cite{snodin2016} (\cref{fig:snodin-diagram}).
This system has 12 staple types that bind to the scaffold, each with two 16-\ac{nt} binding domains.
The simulations were run in under three hours of walltime on a commodity cluster.
To determine the extent of assembly, we look at two order parameters: the number of staples (mis)bound to the scaffold, and the number of bound domain pairs that have formed.
As can be seen in \cref{fig:snodin-melting}(a), at low temperatures the system has the number of (mis)bound staples and the number of bound domain pairs expected in the assembled state.
The error bars are quite narrow, which gives us confidence that the averages have converged.
In \cref{fig:snodin-melting}(b), we show a typical assembled configuration.
Unlike its schematic representation in \cref{fig:snodin-diagram}, the conformation of the scaffold is not planar.
That a typical assembled configuration is not a well-ordered planar state is reasonable because the scaffold is relatively unconstrained by staple crossovers, in part because the crossovers that occur connect only relatively close segments of the scaffold.
Moreover, non-planar configurations were also found to be typical of the assembled state of the same system in oxDNA simulations~\cite{snodin2016}.

\begin{figure}[t]
    \includegraphics{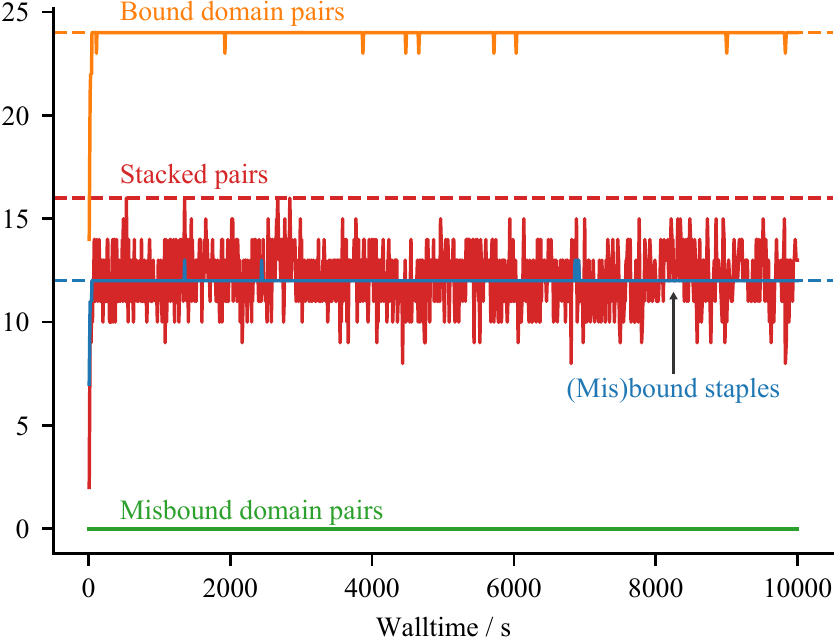}
    \caption{
        \label{fig:snodin-timeseries}
        Order parameter time series for a \SI{330}{\kelvin} replica of a \ac{REMC} simulation of a 24-binding-domain scaffold.
        The simulations were run on a single node of a commodity cluster.
        The dashed lines correspond to the expected order parameter values in the assembled or fully stacked assembled configurations.
    }
\end{figure}

We can examine the extent of the assembled state's structural disorder by looking at the number of stacked binding domain pairs.
In the assembled state, the planar configuration is also the configuration that maximizes the number of stacked binding domain pairs.
As can be seen in \cref{fig:snodin-melting}, the average value of this order parameter converges to a value that is well below the fully stacked assembled configuration at the lower temperatures.
Nevertheless, an examination of a time series of the number of stacked binding domain pairs (\cref{fig:snodin-timeseries}) reveals that, while the average number of stacked binding domain pairs is below that of a fully stacked assembled system, the simulation does sample such configurations.
The fact that the simulations generate such configurations and the large degree of fluctuation in the number of stacked domain pairs gives us confidence that the simulation methods are able to sample origami configurations effectively even in near- and fully assembled states.
Simulations generally result in full assembly within about one hundred seconds of walltime, and fully stacked assembled configurations within an hour.
To give an idea of how the simulations proceed, we have compiled a short video of a serial simulation of the assembly at \SI{335}{\kelvin} (see \ac{SI}).

\begin{figure}[t]
    \includegraphics{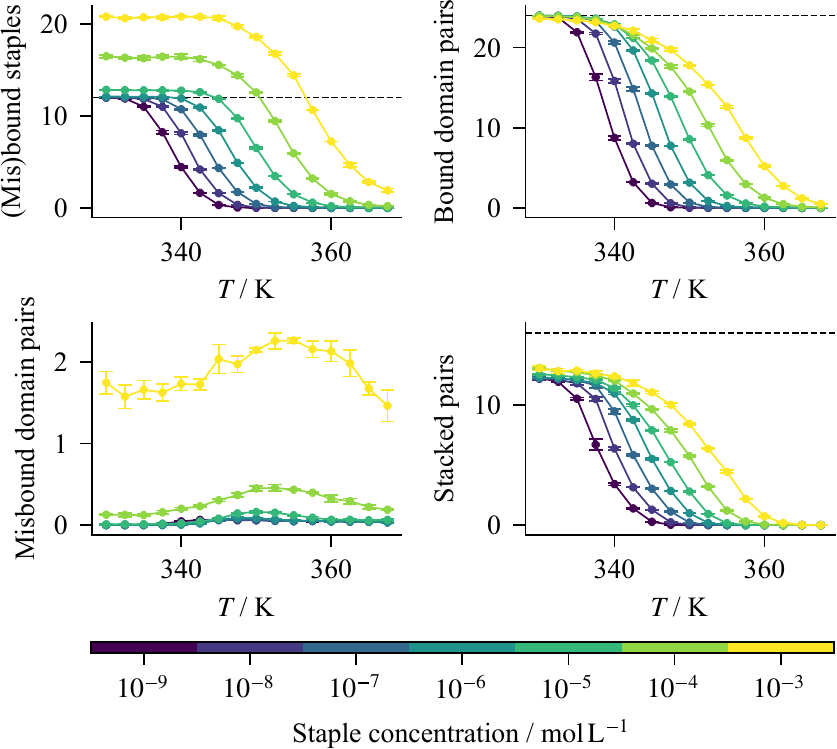}
    \caption{
        \label{fig:staples-melt}
        Mean order parameters plotted against temperature for a range of staple concentrations.
        The dashed lines correspond to the expected order parameter values in the assembled or fully stacked assembled configurations.
        The error bars represent the standard error in the means across three independent simulations.
        Simulations were run with a monovalent cation concentration of \SI{0.5}{\Molar} and a stacking energy of $-1000\,k_\text{B}\si{\kelvin}$.
    }
\end{figure}

The efficiency of our model and sampling methods allows us to run simulations across a range of assembly conditions and design parameters.
While \SI{100}{\nano\Molar} is a typical value for staple concentrations, the concentration used for a particular assembly protocol commonly varies from tens to hundreds of \si{\nano\Molar}.
To see how staple concentration affects the assembly of this system within and beyond the ranges found in experimental conditions, we ran simulations with staple concentrations from \SI{1}{\nano\Molar} to \SI{1}{\milli\Molar} in intervals of factors of 10.
As can be seen in \cref{fig:staples-melt}, at low temperatures, from \SI{1}{\nano\Molar} to \SI{1}{\micro\Molar}, the order parameters indicate that the assembled state is the prevalent structure, with the melting temperature shifting to higher values as the concentration is increased.
However, at \SI{10}{\micro\Molar}, the average number of (mis)bound staples exceeds 12 and the number of misbound domain pairs is near zero, indicating that at least some of the configurations now have two of the same type of staple (mis)bound to the scaffold.
This situation is referred to as `blocking'~\cite{snodin2016} because such staples prevent each other from fully binding to the scaffold.
This is also approximately the staple concentration used in the simulations of Snodin \textit{et al.}~\cite{snodin2016}, who speculated that they seemed to be in a range in which blocking was somewhat favourable.
Blocking becomes substantially more prominent at \SI{100}{\micro\Molar}, at which there are now significant contributions from configurations that have blocked staples for more than one staple type.
The number of stacked binding domain pairs also significantly increases at such high staple concentrations.
The reason for this behaviour is that with multiple staples of the same type bound to the system, there will be fewer crossovers, which can allow for longer segments of stacked helices.
Also at \SI{100}{\micro\Molar}, misbinding begins to play a significant role, and becomes even more substantial in the \si{\milli\Molar} regime.

\begin{figure}
    \includegraphics{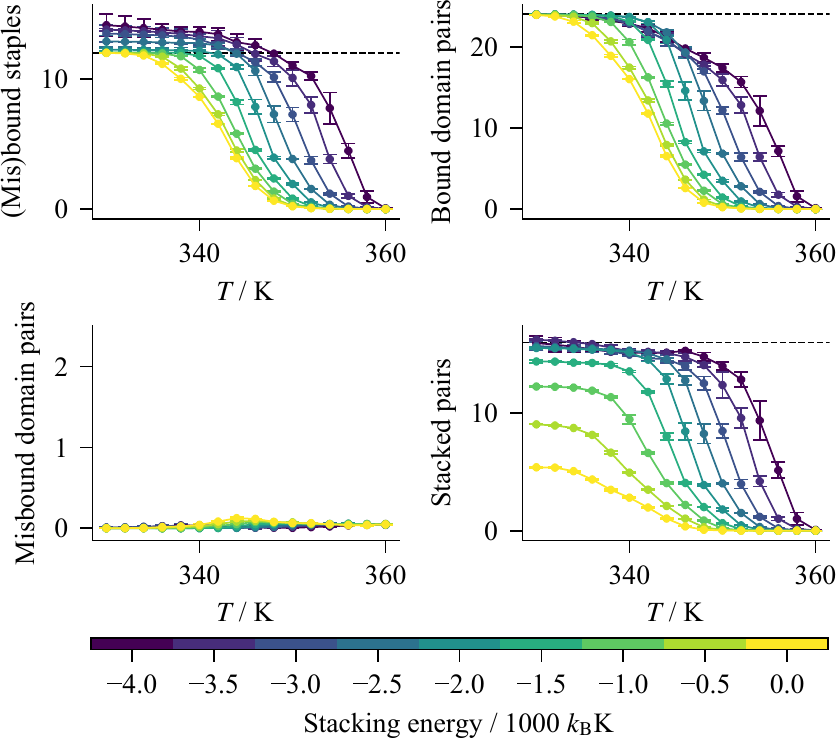}
    \caption{
        \label{fig:stack-melt}
        Mean order parameters plotted against temperature for a range of stacking energies.
        The dashed lines correspond to the expected order parameter values in the assembled or fully stacked assembled configurations.
        The error bars represent the standard error in the means across three independent simulations.
        Simulations were run with a staple concentration of \SI{100}{\nano\Molar} and a monovalent cation concentration of \SI{0.5}{\Molar}.
    }
\end{figure}

Because our choice of the stacking energy was somewhat crudely determined, we ran further simulations with a range of stacking energies to see how strong an effect the choice can have on the thermodynamic assembly behaviour.
According to \cref{fig:stack-melt}, the average number of stacked binding domain pairs changes quite dramatically when the stacking energy is halved or doubled.
The melting temperature is also shifted, although not as dramatically.
When the stacking energy is doubled, the number of stacked binding domain pairs plateaus at nearly the value expected in the planar state.
However, beyond this, the average number of (mis)bound staples exceeds that expected in the assembled state.
It seems that if the stacking energy is sufficiently favourable, the sequence specificity of hybridization can become overshadowed by the non-specific stacking energy.
While our choice of stacking energy is in the right range, because of the sensitivity of the average number of stacked binding domain pairs to this value, if we want to make a more direct comparison between our model and real experiments, the stacking energy should be tuned such that the ratio of planar to non-planar configurations matches experimental values.

While salt concentration can also play a role in the self-assembly behaviour, its primary effect is to shift the melting temperature slightly (see \cref{fig:cation-melt}).
We have only included monovalent cation dependence in our version of the \ac{NN} model, but non-monovalent cations, particularly Mg$^{2+}$, are commonly used in experimental set-ups.
Such ions can be accounted for in a crude manner by simply increasing the effective monovalent cation concentration.
It is possible to use more general corrections to account for such ions within the \ac{NN} model; however, since the effect is relatively small given our model's intended level of accuracy, we have not included such corrections in our model at present.

\begin{figure}
    \includegraphics{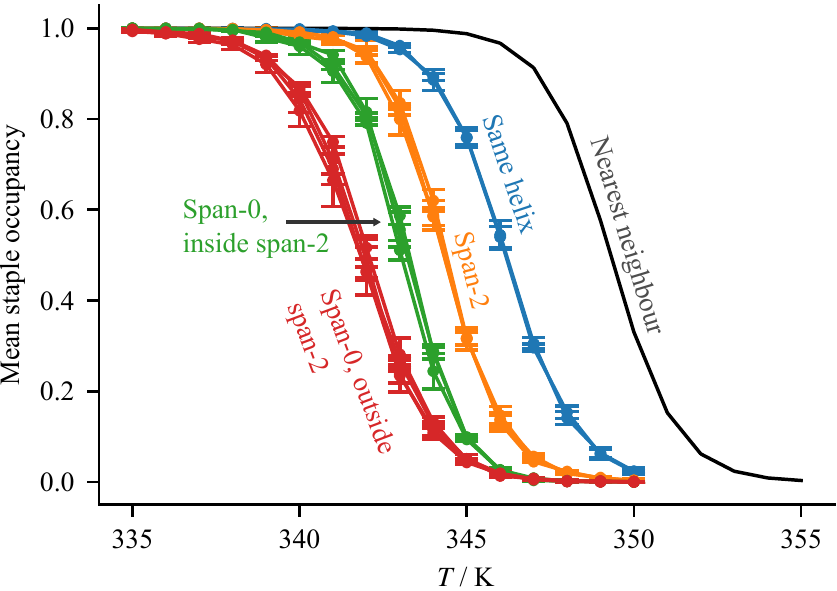}
    \caption{
        \label{fig:staple-curves}
        Mean staple occupancy plotted against temperature for simulations with averaged hybridization free energies.
        The \ac{NN} values were calculated directly with the averaged hybridization free energies by assuming the staple strands are in excess of the scaffold strands.
        The error bars represent the standard error in the means across three independent simulations.
        Simulations were run with the same parameters as the simulations referred to in \cref{fig:snodin-melting}.
    }
\end{figure}

In order to see how the thermodynamic behaviour of individual staples is affected by the scaffold, we ran simulations of the same system, but with the sequence specific hybridization free energies replaced by their average value.
We computed two different averages: one over all the bound pairs and one over all the misbound pairs.
In \cref{fig:staple-curves}, the mean staple occupancy curves are plotted for all staples in the system.
In general, staples with two binding domains can be classed by the number of scaffold binding domains that are spanned by the staple binding domains in the assembled structure.
In the target structure we are assembling here, there are those that span zero and two scaffold binding domains, as well as those that do not have a crossover at all.

The curves of the individual staples turn out to be grouped by these structural classifications.
Those with the highest melting temperatures are those that have no crossovers (same helix; see \cref{fig:snodin-diagram}(b), staples 1 and 12), followed by those that span two scaffold binding domains (span-2; see \cref{fig:snodin-diagram}(b), staples 3, 6, and 9), followed by those that span no scaffold binding domains (span-0; see \cref{fig:snodin-diagram}(b), staples 2, 4, 5, 7, 8, 10, and 11).
The staples with no crossovers are expected to be the most stable, as they have an extra stacking interaction between the two domains compared to the staples that have a crossover.
Further, these staples happen to occur at the termini of the scaffold, so the rigidity that they introduce is placed at a point that will restrict the configuration of the scaffold the least.
However, these staples are still shifted to lower melting temperatures relative to the pure \ac{NN} curve.

The staples that span no scaffold domains involve two crossovers, one with the staple strand, and one with the scaffold strand.
A double crossover will restrict the configuration of the scaffold more than a single crossover, so it is expected that these double crossover bound domains pairs will have a lower melting temperature than those with just one strand crossover.
The curves of the staples involving double crossovers are further split into two distinct groups.
The staples with higher melting temperatures turn out to be those that are within the span of a staple that spans two scaffold binding domains (span-0, inside span-2; see \cref{fig:snodin-diagram}(b), staples 4, 7, and 10), while the staples with the lower melting temperature are those that are not within the span of any other staples (span-0, outside span-2; see \cref{fig:snodin-diagram}(b), staples 2, 5, 8, and 11).
Again, this seems reasonable because the staples that span two scaffold domains will already restrict the scaffold, such that there is a smaller entropic penalty for the staples within their span.

This analysis suggests another possible way of selecting the stacking energy: we could choose a value at which the mean staple occupancy curve of a two-binding-domain helix with no breaks in the backbone overlaps with the \ac{NN} mean staple occupancy curve.
Simulations of such a system reveal that the stacking energy would need to be approximately double that of the value that we selected via the comparison of stacking free-energy differences between our model and experiment.
However, the number of stacked binding domain pairs for a system with such a favourable stacking energy (\cref{fig:stack-melt}) suggests that this would make the system on average nearly planar, which contradicts the simulations of Snodin \textit{et al.}~\cite{snodin2016}.
This suggests that the entropy differences between pairs of bound-domain pairs with an intact backbone and pairs without are not as large as they should be, and so it may be that one stacking term should be used for staples that bind to two contiguous scaffold binding domains, and another for all other pairs.
Nevertheless, for the level of accuracy this model is designed for, it may be sufficient to choose a single stacking energy that is optimal for staples that are involved in crossovers, as staples that bind contiguously to the scaffold at multiple binding domains to form a single helix are uncommon.

\begin{figure}[t]
    \includegraphics{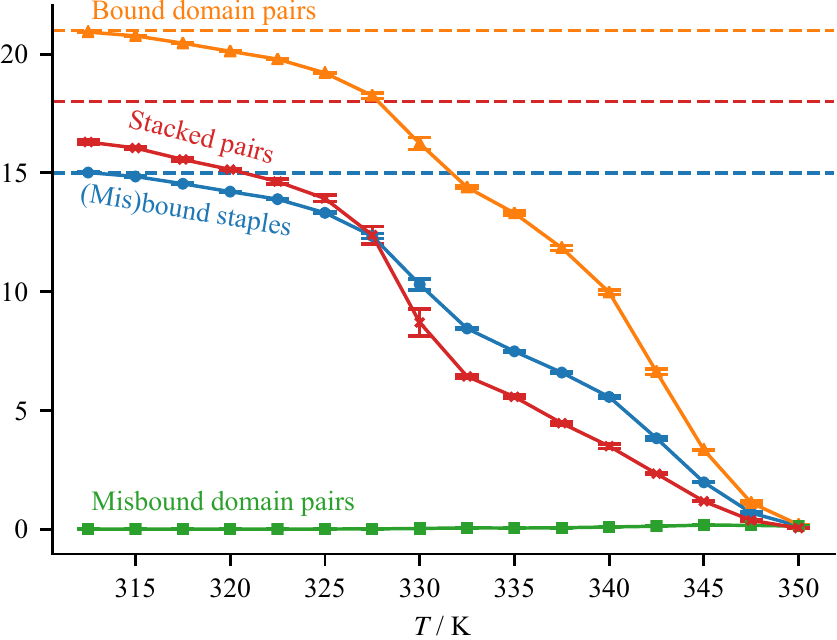}
    \caption{
        \label{fig:halftile-melting}
        Mean order parameters plotted against temperature for a 21-binding-domain scaffold.
        The dashed lines correspond to the expected order parameter values in the assembled or fully stacked assembled configurations.
        The error bars represent the standard error in the means across three independent simulations.
        Simulations were run with the same parameters as the simulations referred to in \cref{fig:snodin-melting}.
    }
\end{figure}

The 24-binding-domain scaffold system contains staples that span at most two scaffold binding domains, which are relatively short spans compared to typical origami structures.
To test whether our sampling methods are able to handle a system with staples that span longer regions of the scaffold, we also ran simulations of a 21-binding-domain scaffold system with staples that span 0, 2, 4, 6, 8, 10 and 12 scaffold binding domains (\cref{fig:halftile-overview}).
The fully stacked assembled state involves three parallel helices composed of seven binding domains each; the design is a subset of the rectangular tile design presented in the original DNA origami paper~\cite{rothemund2006} (the top three rows on the left side of the seam).
The assembly of this system is further complicated by the presence of single-domain staples, which are expected not to bind until significantly lower temperatures than the two-binding-domain staples.
The simulations were again run for under three hours of walltime on a commodity cluster.
The relevant order parameters as a function of temperature are plotted in \cref{fig:halftile-melting}.
As with the 24-binding-domain system, at low temperatures, the system is assembled but not fully stacked.
The order parameter curves now display two distinct regions and do not approach the assembled state values until significantly lower temperatures, as expected.

%    \newpage
    % conclusions.tex

\section{Conclusions}
\label{sec:conclusions}

We have introduced a model and sampling methods for simulating DNA origami self-assembly that is computationally feasible, yet includes the structural information most relevant to the assembly process.
We demonstrated that small origamis can be sampled efficiently enough to achieve good statistics for not only one particular set of assembly conditions and design parameters, but for a range of values of these variables.
It is difficult to predict how the approach will scale with system size, as we expect this may be highly dependent on the specifics of the origami design. 
However, even if simulating the self-assembly of very large systems may not yet be tractable, much insight can be gained from studying smaller origami system.

For example, we can use the model to study thermodynamic properties of origami designs, such as the relative stability of staples, the types and degree of staple binding cooperativity, or the effects of scaffold routing and loop closure on the cost of staple binding.
Because we use Monte Carlo simulations to sample configuration space, we cannot directly study dynamical quantities.
However, we can calculate free-energy barriers along selected order parameters, which in turn could be used to estimate relative rates between different assembly pathways.
Such calculations would allow us to pursue questions relating to the kinetics of assembly, such as whether there is a nucleation barrier, and how it depends on assembly conditions and staple design.
We may also be able to shed some light on whether and why hysteresis occurs for a given design and set of assembly conditions.

There are several caveats to our approach.
We assume that the staples are always in excess of the scaffold.
If that were not the case, the assumption that the free staple concentration remains constant regardless of the degree of assembly would become less convincing.
One solution may be to reduce the free staple concentration relative to the total staple concentration based on the average number of staples (mis)bound to the scaffold.
Of course, because simulations must be run to determine the average staple occupancy on the scaffold, this would require an initial guess and subsequent iterations to converge to a consistent value.
An alternative solution may be to make the free staple concentration a function of the number of staples currently bound to the system.
While not ideal, it may be sufficient for the level of accuracy the model is intended to provide.

It has been found experimentally that the stacking free energy is sequence specific~\cite{protozanova2004,*yakovchuk2006} and depends on both temperature and salt concentration.
It has been observed to range from below \SI{-10}{\kilo\joule\per\mole} to slightly above \SI{1}{\kilo\joule\per\mole} (i.e.~for some sequences and conditions, stacking is slightly disfavoured), which corresponds to a stacking energy range in our model of half to double the chosen value.
As discussed in the results section, the mean number of stacked binding domain pairs shifts substantially over this range of stacking energies.
While some of the temperature dependence is taken into account here by the explicit modeling of some of the entropic contribution to the stacking free energy, the sequence specificity and salt dependence is not accounted for.
For this pilot study, a roughly selected constant value is sufficient to demonstrate that the model is reasonable, but in future studies, we may also consider using sequence-specific salt-dependent stacking energies for more accurate predictions for a particular design.

Here, we have only considered binding domains that are 16~\ac{nt} in length.
Because of the level of flexibility we assume at the crossovers, it is possible for some fully stacked configurations with only 16~\ac{nt} to be non-planar.
However, to model structures that are intended to be assembled into explicitly 3D structures, it will be necessary to extend the model to other lengths of binding domains.
This will primarily entail constructing a stacking potential for each binding domain length considered.
We intend to pursue this in future work.

Finally, we have used simple arguments to support our choices of which kinked configurations are to be allowed and which are to be disallowed within our model.
It may well be that different choices could improve both the reproduction of the balance of the energy/entropy trade-off of stacked assembled configurations and the structural accuracy of the model.
Using a more detailed DNA model such as oxDNA, one could run simulations of helices with breaks in the backbone or simulations of helices with crossovers in order to provide a more detailed reference point from which to determine which configurations are sensible to allow.
Furthermore, another term could be introduced into the Hamiltonian to weight kinked configurations based on their frequency in the higher resolution simulations.
However, because these improvements would only be feasibly applicable to at most four-body interactions, it would still be possible to draw allowed and even energetically favourable model configurations whose physical interpretations are in fact non-physical (see discussion in \cref{sec:appendix-kinks}).
If an even more accurate model were desired, another vector could be introduced to each binding domain that would explicitly represent the helical axis.
Such a model would allow us to control more finely the level of flexibility afforded to kinked segments in the structure and would make it more straightforward to prevent some of the non-physical configurations without introducing further many-body interactions.

While such modifications may improve the accuracy of the model, they would also be costly in both development time and simulation time, and we do not expect they would fundamentally alter the results, but rather may incrementally improve them.
For studying fundamental aspects of DNA origami self-assembly, we believe that such expensive incremental improvements are likely to be of marginal use.
We are therefore hopeful that the use of our model will be able to yield both fundamental and practical insights into the thermodynamics and kinetics of DNA origami self-assembly.

    % acks.tex

\section{Acknowledgements}
\label{sec:acks}

This work was supported by the European Union's Horizon 2020 research and innovation programme under the Marie Sk\l{}odowska-Curie grant agreement No.~642774 (ETN-COLLDENSE).
D.F. acknowledges support from a UKIERI grant: DST-UKIERI-2016-17-0190.

%    \newpage
%    \input{main.bbl}
    \bibliography{main}

%merlin.mbs aipnum4-1.bst 2010-07-25 4.21a (PWD, AO, DPC) hacked
%Control: key (0)
%Control: author (8) initials jnrlst
%Control: editor formatted (1) identically to author
%Control: production of article title (0) allowed
%Control: page (0) single
%Control: year (1) truncated
%Control: production of eprint (-1) disabled
\begin{thebibliography}{45}%
\makeatletter
\providecommand \@ifxundefined [1]{%
 \@ifx{#1\undefined}
}%
\providecommand \@ifnum [1]{%
 \ifnum #1\expandafter \@firstoftwo
 \else \expandafter \@secondoftwo
 \fi
}%
\providecommand \@ifx [1]{%
 \ifx #1\expandafter \@firstoftwo
 \else \expandafter \@secondoftwo
 \fi
}%
\providecommand \natexlab [1]{#1}%
\providecommand \enquote  [1]{``#1''}%
\providecommand \bibnamefont  [1]{#1}%
\providecommand \bibfnamefont [1]{#1}%
\providecommand \citenamefont [1]{#1}%
\providecommand \href@noop [0]{\@secondoftwo}%
\providecommand \href [0]{\begingroup \@sanitize@url \@href}%
\providecommand \@href[1]{\@@startlink{#1}\@@href}%
\providecommand \@@href[1]{\endgroup#1\@@endlink}%
\providecommand \@sanitize@url [0]{\catcode `\\12\catcode `\$12\catcode
  `\&12\catcode `\#12\catcode `\^12\catcode `\_12\catcode `\%12\relax}%
\providecommand \@@startlink[1]{}%
\providecommand \@@endlink[0]{}%
\providecommand \url  [0]{\begingroup\@sanitize@url \@url }%
\providecommand \@url [1]{\endgroup\@href {#1}{\urlprefix }}%
\providecommand \urlprefix  [0]{URL }%
\providecommand \Eprint [0]{\href }%
\providecommand \doibase [0]{http://dx.doi.org/}%
\providecommand \selectlanguage [0]{\@gobble}%
\providecommand \bibinfo  [0]{\@secondoftwo}%
\providecommand \bibfield  [0]{\@secondoftwo}%
\providecommand \translation [1]{[#1]}%
\providecommand \BibitemOpen [0]{}%
\providecommand \bibitemStop [0]{}%
\providecommand \bibitemNoStop [0]{.\EOS\space}%
\providecommand \EOS [0]{\spacefactor3000\relax}%
\providecommand \BibitemShut  [1]{\csname bibitem#1\endcsname}%
\let\auto@bib@innerbib\@empty
%</preamble>
\bibitem [{\citenamefont {Seeman}(1982)}]{seeman1982}%
  \BibitemOpen
  \bibfield  {author} {\bibinfo {author} {\bibfnamefont {N.~C.}\ \bibnamefont
  {Seeman}},\ }\bibfield  {title} {\enquote {\bibinfo {title} {Nucleic acid
  junctions and lattices},}\ }\href {\doibase 10.1016/0022-5193(82)90002-9}
  {\bibfield  {journal} {\bibinfo  {journal} {J. Theor. Biol.}\ }\textbf
  {\bibinfo {volume} {99}},\ \bibinfo {pages} {237 } (\bibinfo {year}
  {1982})}\BibitemShut {NoStop}%
\bibitem [{\citenamefont {Rothemund}(2006)}]{rothemund2006}%
  \BibitemOpen
  \bibfield  {author} {\bibinfo {author} {\bibfnamefont {P.}~\bibnamefont
  {Rothemund}},\ }\bibfield  {title} {\enquote {\bibinfo {title} {Folding {DNA}
  to create nanoscale shapes and patterns},}\ }\href {\doibase
  10.1038/nature04586} {\bibfield  {journal} {\bibinfo  {journal} {Nature}\
  }\textbf {\bibinfo {volume} {440}},\ \bibinfo {pages} {297} (\bibinfo {year}
  {2006})}\BibitemShut {NoStop}%
\bibitem [{\citenamefont {Douglas}\ \emph {et~al.}(2009)\citenamefont
  {Douglas}, \citenamefont {Dietz}, \citenamefont {Liedl}, \citenamefont
  {Hoegberg}, \citenamefont {Graf},\ and\ \citenamefont {Shih}}]{douglas2009}%
  \BibitemOpen
  \bibfield  {author} {\bibinfo {author} {\bibfnamefont {S.~M.}\ \bibnamefont
  {Douglas}}, \bibinfo {author} {\bibfnamefont {H.}~\bibnamefont {Dietz}},
  \bibinfo {author} {\bibfnamefont {T.}~\bibnamefont {Liedl}}, \bibinfo
  {author} {\bibfnamefont {B.}~\bibnamefont {Hoegberg}}, \bibinfo {author}
  {\bibfnamefont {F.}~\bibnamefont {Graf}}, \ and\ \bibinfo {author}
  {\bibfnamefont {W.~M.}\ \bibnamefont {Shih}},\ }\bibfield  {title} {\enquote
  {\bibinfo {title} {Self-assembly of {DNA} into nanoscale three-dimensional
  shapes},}\ }\href {\doibase 10.1038/nature08016} {\bibfield  {journal}
  {\bibinfo  {journal} {Nature}\ }\textbf {\bibinfo {volume} {459}},\ \bibinfo
  {pages} {414} (\bibinfo {year} {2009})}\BibitemShut {NoStop}%
\bibitem [{\citenamefont {Castro}\ \emph {et~al.}(2011)\citenamefont {Castro},
  \citenamefont {Kilchherr}, \citenamefont {Kim}, \citenamefont {Shiao},
  \citenamefont {Wauer}, \citenamefont {Wortmann}, \citenamefont {Bathe},\ and\
  \citenamefont {Dietz}}]{castro2011}%
  \BibitemOpen
  \bibfield  {author} {\bibinfo {author} {\bibfnamefont {C.~E.}\ \bibnamefont
  {Castro}}, \bibinfo {author} {\bibfnamefont {F.}~\bibnamefont {Kilchherr}},
  \bibinfo {author} {\bibfnamefont {D.-N.}\ \bibnamefont {Kim}}, \bibinfo
  {author} {\bibfnamefont {E.~L.}\ \bibnamefont {Shiao}}, \bibinfo {author}
  {\bibfnamefont {T.}~\bibnamefont {Wauer}}, \bibinfo {author} {\bibfnamefont
  {P.}~\bibnamefont {Wortmann}}, \bibinfo {author} {\bibfnamefont
  {M.}~\bibnamefont {Bathe}}, \ and\ \bibinfo {author} {\bibfnamefont
  {H.}~\bibnamefont {Dietz}},\ }\bibfield  {title} {\enquote {\bibinfo {title}
  {A primer to scaffolded {DNA} origami},}\ }\href {\doibase
  10.1038/NMETH.1570} {\bibfield  {journal} {\bibinfo  {journal} {Nat.
  Methods}\ }\textbf {\bibinfo {volume} {8}},\ \bibinfo {pages} {221} (\bibinfo
  {year} {2011})}\BibitemShut {NoStop}%
\bibitem [{\citenamefont {Han}\ \emph {et~al.}(2013)\citenamefont {Han},
  \citenamefont {Pal}, \citenamefont {Yang}, \citenamefont {Jiang},
  \citenamefont {Nangreave}, \citenamefont {Liu},\ and\ \citenamefont
  {Yan}}]{han2013}%
  \BibitemOpen
  \bibfield  {author} {\bibinfo {author} {\bibfnamefont {D.}~\bibnamefont
  {Han}}, \bibinfo {author} {\bibfnamefont {S.}~\bibnamefont {Pal}}, \bibinfo
  {author} {\bibfnamefont {Y.}~\bibnamefont {Yang}}, \bibinfo {author}
  {\bibfnamefont {S.}~\bibnamefont {Jiang}}, \bibinfo {author} {\bibfnamefont
  {J.}~\bibnamefont {Nangreave}}, \bibinfo {author} {\bibfnamefont
  {Y.}~\bibnamefont {Liu}}, \ and\ \bibinfo {author} {\bibfnamefont
  {H.}~\bibnamefont {Yan}},\ }\bibfield  {title} {\enquote {\bibinfo {title}
  {{DNA} gridiron nanostructures based on four-arm junctions},}\ }\href
  {\doibase 10.1126/science.1232252} {\bibfield  {journal} {\bibinfo  {journal}
  {Science}\ }\textbf {\bibinfo {volume} {339}},\ \bibinfo {pages} {1412}
  (\bibinfo {year} {2013})}\BibitemShut {NoStop}%
\bibitem [{\citenamefont {Benson}\ \emph {et~al.}(2015)\citenamefont {Benson},
  \citenamefont {Mohammed}, \citenamefont {Gardell}, \citenamefont {Masich},
  \citenamefont {Czeizler}, \citenamefont {Orponen},\ and\ \citenamefont
  {Hogberg}}]{benson2015}%
  \BibitemOpen
  \bibfield  {author} {\bibinfo {author} {\bibfnamefont {E.}~\bibnamefont
  {Benson}}, \bibinfo {author} {\bibfnamefont {A.}~\bibnamefont {Mohammed}},
  \bibinfo {author} {\bibfnamefont {J.}~\bibnamefont {Gardell}}, \bibinfo
  {author} {\bibfnamefont {S.}~\bibnamefont {Masich}}, \bibinfo {author}
  {\bibfnamefont {E.}~\bibnamefont {Czeizler}}, \bibinfo {author}
  {\bibfnamefont {P.}~\bibnamefont {Orponen}}, \ and\ \bibinfo {author}
  {\bibfnamefont {B.}~\bibnamefont {Hogberg}},\ }\bibfield  {title} {\enquote
  {\bibinfo {title} {{DNA} rendering of polyhedral meshes at the nanoscale},}\
  }\href {\doibase 10.1038/nature14586} {\bibfield  {journal} {\bibinfo
  {journal} {Nature}\ }\textbf {\bibinfo {volume} {523}},\ \bibinfo {pages}
  {441} (\bibinfo {year} {2015})}\BibitemShut {NoStop}%
\bibitem [{\citenamefont {Castro}\ \emph {et~al.}(2015)\citenamefont {Castro},
  \citenamefont {Su}, \citenamefont {Marras}, \citenamefont {Zhou},\ and\
  \citenamefont {Johnson}}]{castro2015}%
  \BibitemOpen
  \bibfield  {author} {\bibinfo {author} {\bibfnamefont {C.~E.}\ \bibnamefont
  {Castro}}, \bibinfo {author} {\bibfnamefont {H.-J.}\ \bibnamefont {Su}},
  \bibinfo {author} {\bibfnamefont {A.~E.}\ \bibnamefont {Marras}}, \bibinfo
  {author} {\bibfnamefont {L.}~\bibnamefont {Zhou}}, \ and\ \bibinfo {author}
  {\bibfnamefont {J.}~\bibnamefont {Johnson}},\ }\bibfield  {title} {\enquote
  {\bibinfo {title} {Mechanical design of {DNA} nanostructures},}\ }\href
  {\doibase 10.1039/C4NR07153K} {\bibfield  {journal} {\bibinfo  {journal}
  {Nanoscale}\ }\textbf {\bibinfo {volume} {7}},\ \bibinfo {pages} {5913}
  (\bibinfo {year} {2015})}\BibitemShut {NoStop}%
\bibitem [{\citenamefont {Zhang}\ \emph {et~al.}(2015)\citenamefont {Zhang},
  \citenamefont {Jiang}, \citenamefont {Wu}, \citenamefont {Li}, \citenamefont
  {Mao}, \citenamefont {Liu},\ and\ \citenamefont {Yan}}]{zhang2015b}%
  \BibitemOpen
  \bibfield  {author} {\bibinfo {author} {\bibfnamefont {F.}~\bibnamefont
  {Zhang}}, \bibinfo {author} {\bibfnamefont {S.}~\bibnamefont {Jiang}},
  \bibinfo {author} {\bibfnamefont {S.}~\bibnamefont {Wu}}, \bibinfo {author}
  {\bibfnamefont {Y.}~\bibnamefont {Li}}, \bibinfo {author} {\bibfnamefont
  {C.}~\bibnamefont {Mao}}, \bibinfo {author} {\bibfnamefont {Y.}~\bibnamefont
  {Liu}}, \ and\ \bibinfo {author} {\bibfnamefont {H.}~\bibnamefont {Yan}},\
  }\bibfield  {title} {\enquote {\bibinfo {title} {Complex wireframe {DNA}
  origami nanostructures with multi-arm junction vertices},}\ }\href {\doibase
  10.1038/NNANO.2015.162} {\bibfield  {journal} {\bibinfo  {journal} {Nat.
  Nanotechnol.}\ }\textbf {\bibinfo {volume} {10}},\ \bibinfo {pages} {779}
  (\bibinfo {year} {2015})}\BibitemShut {NoStop}%
\bibitem [{\citenamefont {Ke}\ \emph {et~al.}(2008)\citenamefont {Ke},
  \citenamefont {Lindsay}, \citenamefont {Chang}, \citenamefont {Liu},\ and\
  \citenamefont {Yan}}]{ke2008}%
  \BibitemOpen
  \bibfield  {author} {\bibinfo {author} {\bibfnamefont {Y.}~\bibnamefont
  {Ke}}, \bibinfo {author} {\bibfnamefont {S.}~\bibnamefont {Lindsay}},
  \bibinfo {author} {\bibfnamefont {Y.}~\bibnamefont {Chang}}, \bibinfo
  {author} {\bibfnamefont {Y.}~\bibnamefont {Liu}}, \ and\ \bibinfo {author}
  {\bibfnamefont {H.}~\bibnamefont {Yan}},\ }\bibfield  {title} {\enquote
  {\bibinfo {title} {Self-assembled water-soluble nucleic acid probe tiles for
  label-free {RNA} hybridization assays},}\ }\href {\doibase
  10.1126/science.1150082} {\bibfield  {journal} {\bibinfo  {journal}
  {Science}\ }\textbf {\bibinfo {volume} {319}},\ \bibinfo {pages} {180}
  (\bibinfo {year} {2008})}\BibitemShut {NoStop}%
\bibitem [{\citenamefont {Linko}, \citenamefont {Eerikainen},\ and\
  \citenamefont {Kostiainen}(2015)}]{linko2015b}%
  \BibitemOpen
  \bibfield  {author} {\bibinfo {author} {\bibfnamefont {V.}~\bibnamefont
  {Linko}}, \bibinfo {author} {\bibfnamefont {M.}~\bibnamefont {Eerikainen}}, \
  and\ \bibinfo {author} {\bibfnamefont {M.~A.}\ \bibnamefont {Kostiainen}},\
  }\bibfield  {title} {\enquote {\bibinfo {title} {A modular {DNA}
  origami-based enzyme cascade nanoreactor},}\ }\href {\doibase
  10.1039/C4CC08472A} {\bibfield  {journal} {\bibinfo  {journal} {Chem.
  Commun.}\ }\textbf {\bibinfo {volume} {51}},\ \bibinfo {pages} {5351}
  (\bibinfo {year} {2015})}\BibitemShut {NoStop}%
\bibitem [{\citenamefont {Liu}\ \emph {et~al.}(2013)\citenamefont {Liu},
  \citenamefont {Fu}, \citenamefont {Hejesen}, \citenamefont {Yang},
  \citenamefont {Woodbury}, \citenamefont {Gothelf}, \citenamefont {Liu},\ and\
  \citenamefont {Yan}}]{liu2013}%
  \BibitemOpen
  \bibfield  {author} {\bibinfo {author} {\bibfnamefont {M.}~\bibnamefont
  {Liu}}, \bibinfo {author} {\bibfnamefont {J.}~\bibnamefont {Fu}}, \bibinfo
  {author} {\bibfnamefont {C.}~\bibnamefont {Hejesen}}, \bibinfo {author}
  {\bibfnamefont {Y.}~\bibnamefont {Yang}}, \bibinfo {author} {\bibfnamefont
  {N.~W.}\ \bibnamefont {Woodbury}}, \bibinfo {author} {\bibfnamefont
  {K.}~\bibnamefont {Gothelf}}, \bibinfo {author} {\bibfnamefont
  {Y.}~\bibnamefont {Liu}}, \ and\ \bibinfo {author} {\bibfnamefont
  {H.}~\bibnamefont {Yan}},\ }\bibfield  {title} {\enquote {\bibinfo {title} {A
  {DNA} tweezer-actuated enzyme nanoreactor},}\ }\href {\doibase
  10.1038/ncomms3127} {\bibfield  {journal} {\bibinfo  {journal} {Nat.
  Commun.}\ }\textbf {\bibinfo {volume} {4}},\ \bibinfo {pages} {2127}
  (\bibinfo {year} {2013})}\BibitemShut {NoStop}%
\bibitem [{\citenamefont {Maune}\ \emph {et~al.}(2010)\citenamefont {Maune},
  \citenamefont {Han}, \citenamefont {Barish}, \citenamefont {Bockrath},
  \citenamefont {Goddard}, \citenamefont {Rothemund},\ and\ \citenamefont
  {Winfree}}]{maune2010}%
  \BibitemOpen
  \bibfield  {author} {\bibinfo {author} {\bibfnamefont {H.~T.}\ \bibnamefont
  {Maune}}, \bibinfo {author} {\bibfnamefont {S.-p.}\ \bibnamefont {Han}},
  \bibinfo {author} {\bibfnamefont {R.~D.}\ \bibnamefont {Barish}}, \bibinfo
  {author} {\bibfnamefont {M.}~\bibnamefont {Bockrath}}, \bibinfo {author}
  {\bibfnamefont {W.~A.}\ \bibnamefont {Goddard}, \bibfnamefont {III}},
  \bibinfo {author} {\bibfnamefont {P.~W.~K.}\ \bibnamefont {Rothemund}}, \
  and\ \bibinfo {author} {\bibfnamefont {E.}~\bibnamefont {Winfree}},\
  }\bibfield  {title} {\enquote {\bibinfo {title} {Self-assembly of carbon
  nanotubes into two-dimensional geometries using {DNA} origami templates},}\
  }\href {\doibase 10.1038/NNANO.2009.311} {\bibfield  {journal} {\bibinfo
  {journal} {Nat. Nanotechnol.}\ }\textbf {\bibinfo {volume} {5}},\ \bibinfo
  {pages} {61} (\bibinfo {year} {2010})}\BibitemShut {NoStop}%
\bibitem [{\citenamefont {Linko}, \citenamefont {Ora},\ and\ \citenamefont
  {Kostiainen}(2015)}]{linko2015}%
  \BibitemOpen
  \bibfield  {author} {\bibinfo {author} {\bibfnamefont {V.}~\bibnamefont
  {Linko}}, \bibinfo {author} {\bibfnamefont {A.}~\bibnamefont {Ora}}, \ and\
  \bibinfo {author} {\bibfnamefont {M.~A.}\ \bibnamefont {Kostiainen}},\
  }\bibfield  {title} {\enquote {\bibinfo {title} {{DNA} nanostructures as
  smart drug-delivery vehicles and molecular devices},}\ }\href {\doibase
  10.1016/j.tibtech.2015.08.001} {\bibfield  {journal} {\bibinfo  {journal}
  {Trends Biotechnol.}\ }\textbf {\bibinfo {volume} {33}},\ \bibinfo {pages}
  {586} (\bibinfo {year} {2015})}\BibitemShut {NoStop}%
\bibitem [{\citenamefont {Cademartiri}\ and\ \citenamefont
  {Bishop}(2015)}]{cademartiri2015}%
  \BibitemOpen
  \bibfield  {author} {\bibinfo {author} {\bibfnamefont {L.}~\bibnamefont
  {Cademartiri}}\ and\ \bibinfo {author} {\bibfnamefont {K.~J.~M.}\
  \bibnamefont {Bishop}},\ }\bibfield  {title} {\enquote {\bibinfo {title}
  {Programmable self-assembly},}\ }\href {\doibase 10.1038/nmat4184} {\bibfield
   {journal} {\bibinfo  {journal} {Nat. Mater.}\ }\textbf {\bibinfo {volume}
  {14}},\ \bibinfo {pages} {2} (\bibinfo {year} {2015})}\BibitemShut {NoStop}%
\bibitem [{\citenamefont {Frenkel}(2015)}]{frenkel2015}%
  \BibitemOpen
  \bibfield  {author} {\bibinfo {author} {\bibfnamefont {D.}~\bibnamefont
  {Frenkel}},\ }\bibfield  {title} {\enquote {\bibinfo {title} {Order through
  entropy},}\ }\href {\doibase 10.1038/nmat4178} {\bibfield  {journal}
  {\bibinfo  {journal} {Nat. Mater.}\ }\textbf {\bibinfo {volume} {14}},\
  \bibinfo {pages} {9} (\bibinfo {year} {2015})}\BibitemShut {NoStop}%
\bibitem [{\citenamefont {Ke}\ \emph {et~al.}(2012)\citenamefont {Ke},
  \citenamefont {Ong}, \citenamefont {Shih},\ and\ \citenamefont
  {Yin}}]{ke2012}%
  \BibitemOpen
  \bibfield  {author} {\bibinfo {author} {\bibfnamefont {Y.}~\bibnamefont
  {Ke}}, \bibinfo {author} {\bibfnamefont {L.~L.}\ \bibnamefont {Ong}},
  \bibinfo {author} {\bibfnamefont {W.~M.}\ \bibnamefont {Shih}}, \ and\
  \bibinfo {author} {\bibfnamefont {P.}~\bibnamefont {Yin}},\ }\bibfield
  {title} {\enquote {\bibinfo {title} {Three-dimensional structures
  self-assembled from {DNA} bricks},}\ }\href {\doibase
  10.1126/science.1227268} {\bibfield  {journal} {\bibinfo  {journal}
  {Science}\ }\textbf {\bibinfo {volume} {338}},\ \bibinfo {pages} {1177}
  (\bibinfo {year} {2012})}\BibitemShut {NoStop}%
\bibitem [{\citenamefont {Sobczak}\ \emph {et~al.}(2012)\citenamefont
  {Sobczak}, \citenamefont {Martin}, \citenamefont {Gerling},\ and\
  \citenamefont {Dietz}}]{sobczak2012}%
  \BibitemOpen
  \bibfield  {author} {\bibinfo {author} {\bibfnamefont {J.-P.~J.}\
  \bibnamefont {Sobczak}}, \bibinfo {author} {\bibfnamefont {T.~G.}\
  \bibnamefont {Martin}}, \bibinfo {author} {\bibfnamefont {T.}~\bibnamefont
  {Gerling}}, \ and\ \bibinfo {author} {\bibfnamefont {H.}~\bibnamefont
  {Dietz}},\ }\bibfield  {title} {\enquote {\bibinfo {title} {Rapid folding of
  {DNA} into nanoscale shapes at constant temperature},}\ }\href {\doibase
  10.1126/science.1229919} {\bibfield  {journal} {\bibinfo  {journal}
  {Science}\ }\textbf {\bibinfo {volume} {338}},\ \bibinfo {pages} {1458}
  (\bibinfo {year} {2012})}\BibitemShut {NoStop}%
\bibitem [{\citenamefont {Dunn}\ \emph {et~al.}(2015)\citenamefont {Dunn},
  \citenamefont {Dannenberg}, \citenamefont {Ouldridge}, \citenamefont
  {Kwiatkowska}, \citenamefont {Turberfield},\ and\ \citenamefont
  {Bath}}]{dunn2015}%
  \BibitemOpen
  \bibfield  {author} {\bibinfo {author} {\bibfnamefont {K.~E.}\ \bibnamefont
  {Dunn}}, \bibinfo {author} {\bibfnamefont {F.}~\bibnamefont {Dannenberg}},
  \bibinfo {author} {\bibfnamefont {T.~E.}\ \bibnamefont {Ouldridge}}, \bibinfo
  {author} {\bibfnamefont {M.}~\bibnamefont {Kwiatkowska}}, \bibinfo {author}
  {\bibfnamefont {A.~J.}\ \bibnamefont {Turberfield}}, \ and\ \bibinfo {author}
  {\bibfnamefont {J.}~\bibnamefont {Bath}},\ }\bibfield  {title} {\enquote
  {\bibinfo {title} {Guiding the folding pathway of {DNA} origami},}\ }\href
  {\doibase 10.1038/nature14860} {\bibfield  {journal} {\bibinfo  {journal}
  {Nature}\ }\textbf {\bibinfo {volume} {525}},\ \bibinfo {pages} {82}
  (\bibinfo {year} {2015})}\BibitemShut {NoStop}%
\bibitem [{\citenamefont {Dannenberg}\ \emph {et~al.}(2015)\citenamefont
  {Dannenberg}, \citenamefont {Dunn}, \citenamefont {Bath}, \citenamefont
  {Kwiatkowska}, \citenamefont {Turberfield},\ and\ \citenamefont
  {Ouldridge}}]{dannenberg2015}%
  \BibitemOpen
  \bibfield  {author} {\bibinfo {author} {\bibfnamefont {F.}~\bibnamefont
  {Dannenberg}}, \bibinfo {author} {\bibfnamefont {K.~E.}\ \bibnamefont
  {Dunn}}, \bibinfo {author} {\bibfnamefont {J.}~\bibnamefont {Bath}}, \bibinfo
  {author} {\bibfnamefont {M.}~\bibnamefont {Kwiatkowska}}, \bibinfo {author}
  {\bibfnamefont {A.~J.}\ \bibnamefont {Turberfield}}, \ and\ \bibinfo {author}
  {\bibfnamefont {T.~E.}\ \bibnamefont {Ouldridge}},\ }\bibfield  {title}
  {\enquote {\bibinfo {title} {Modelling {DNA} origami self-assembly at the
  domain level},}\ }\href {\doibase 10.1063/1.4933426} {\bibfield  {journal}
  {\bibinfo  {journal} {J. Chem. Phys.}\ }\textbf {\bibinfo {volume} {143}},\
  (\bibinfo {year} {2015})}\BibitemShut {NoStop}%
\bibitem [{\citenamefont {Wei}\ \emph {et~al.}(2013)\citenamefont {Wei},
  \citenamefont {Nangreave}, \citenamefont {Jiang}, \citenamefont {Yan},\ and\
  \citenamefont {Liu}}]{wei2013}%
  \BibitemOpen
  \bibfield  {author} {\bibinfo {author} {\bibfnamefont {X.}~\bibnamefont
  {Wei}}, \bibinfo {author} {\bibfnamefont {J.}~\bibnamefont {Nangreave}},
  \bibinfo {author} {\bibfnamefont {S.}~\bibnamefont {Jiang}}, \bibinfo
  {author} {\bibfnamefont {H.}~\bibnamefont {Yan}}, \ and\ \bibinfo {author}
  {\bibfnamefont {Y.}~\bibnamefont {Liu}},\ }\bibfield  {title} {\enquote
  {\bibinfo {title} {Mapping the thermal behavior of {DNA} origami
  nanostructures},}\ }\href {\doibase 10.1021/ja4000728} {\bibfield  {journal}
  {\bibinfo  {journal} {J. Am. Chem. Soc.}\ }\textbf {\bibinfo {volume}
  {135}},\ \bibinfo {pages} {6165} (\bibinfo {year} {2013})}\BibitemShut
  {NoStop}%
\bibitem [{\citenamefont {Arbona}, \citenamefont {Aimé},\ and\ \citenamefont
  {Elezgaray}(2013)}]{arbona2013}%
  \BibitemOpen
  \bibfield  {author} {\bibinfo {author} {\bibfnamefont {J.-M.}\ \bibnamefont
  {Arbona}}, \bibinfo {author} {\bibfnamefont {J.-P.}\ \bibnamefont {Aimé}}, \
  and\ \bibinfo {author} {\bibfnamefont {J.}~\bibnamefont {Elezgaray}},\
  }\bibfield  {title} {\enquote {\bibinfo {title} {Cooperativity in the
  annealing of {DNA} origamis},}\ }\href {\doibase 10.1063/1.4773405}
  {\bibfield  {journal} {\bibinfo  {journal} {J. Chem. Phys.}\ }\textbf
  {\bibinfo {volume} {138}},\  (\bibinfo {year} {2013})}\BibitemShut {NoStop}%
\bibitem [{\citenamefont {Arbona}, \citenamefont {Elezgaray},\ and\
  \citenamefont {Aimé}(2012)}]{arbona2012}%
  \BibitemOpen
  \bibfield  {author} {\bibinfo {author} {\bibfnamefont {J.-M.}\ \bibnamefont
  {Arbona}}, \bibinfo {author} {\bibfnamefont {J.}~\bibnamefont {Elezgaray}}, \
  and\ \bibinfo {author} {\bibfnamefont {J.-P.}\ \bibnamefont {Aimé}},\
  }\bibfield  {title} {\enquote {\bibinfo {title} {Modelling the folding of
  {DNA} origami},}\ }\href {\doibase 10.1209/0295-5075/100/28006} {\bibfield
  {journal} {\bibinfo  {journal} {Europhys. Lett.}\ }\textbf {\bibinfo {volume}
  {100}},\ \bibinfo {pages} {28006} (\bibinfo {year} {2012})}\BibitemShut
  {NoStop}%
\bibitem [{\citenamefont {Arbona}, \citenamefont {Aimé},\ and\ \citenamefont
  {Elezgaray}(2012)}]{arbona2012b}%
  \BibitemOpen
  \bibfield  {author} {\bibinfo {author} {\bibfnamefont {J.-M.}\ \bibnamefont
  {Arbona}}, \bibinfo {author} {\bibfnamefont {J.-P.}\ \bibnamefont {Aimé}}, \
  and\ \bibinfo {author} {\bibfnamefont {J.}~\bibnamefont {Elezgaray}},\
  }\bibfield  {title} {\enquote {\bibinfo {title} {Folding of {DNA}
  origamis},}\ }\href {\doibase 10.1080/21553769.2013.768556} {\bibfield
  {journal} {\bibinfo  {journal} {Front. Life Sci.}\ }\textbf {\bibinfo
  {volume} {6}},\ \bibinfo {pages} {11} (\bibinfo {year} {2012})}\BibitemShut
  {NoStop}%
\bibitem [{\citenamefont {Wah}\ \emph {et~al.}(2016)\citenamefont {Wah},
  \citenamefont {David}, \citenamefont {Rudiuk}, \citenamefont {Baigl},\ and\
  \citenamefont {Estevez-Torres}}]{wah2016}%
  \BibitemOpen
  \bibfield  {author} {\bibinfo {author} {\bibfnamefont {J.~L.~T.}\
  \bibnamefont {Wah}}, \bibinfo {author} {\bibfnamefont {C.}~\bibnamefont
  {David}}, \bibinfo {author} {\bibfnamefont {S.}~\bibnamefont {Rudiuk}},
  \bibinfo {author} {\bibfnamefont {D.}~\bibnamefont {Baigl}}, \ and\ \bibinfo
  {author} {\bibfnamefont {A.}~\bibnamefont {Estevez-Torres}},\ }\bibfield
  {title} {\enquote {\bibinfo {title} {Observing and controlling the folding
  pathway of {DNA} origami at the nanoscale},}\ }\href {\doibase
  10.1021/acsnano.5b05972} {\bibfield  {journal} {\bibinfo  {journal} {ACS
  Nano}\ }\textbf {\bibinfo {volume} {10}},\ \bibinfo {pages} {1978} (\bibinfo
  {year} {2016})}\BibitemShut {NoStop}%
\bibitem [{\citenamefont {SantaLucia~Jr.}\ and\ \citenamefont
  {Hicks}(2004)}]{santalucia2004}%
  \BibitemOpen
  \bibfield  {author} {\bibinfo {author} {\bibfnamefont {J.}~\bibnamefont
  {SantaLucia~Jr.}}\ and\ \bibinfo {author} {\bibfnamefont {D.}~\bibnamefont
  {Hicks}},\ }\bibfield  {title} {\enquote {\bibinfo {title} {The
  thermodynamics of {DNA} structural motifs},}\ }\href {\doibase
  10.1146/annurev.biophys.32.110601.141800} {\bibfield  {journal} {\bibinfo
  {journal} {Annu. Rev. Biophys. Biomol. Struct.}\ }\textbf {\bibinfo {volume}
  {33}},\ \bibinfo {pages} {415} (\bibinfo {year} {2004})}\BibitemShut
  {NoStop}%
\bibitem [{\citenamefont {Snodin}\ \emph {et~al.}(2016)\citenamefont {Snodin},
  \citenamefont {Romano}, \citenamefont {Rovigatti}, \citenamefont {Ouldridge},
  \citenamefont {Louis},\ and\ \citenamefont {Doye}}]{snodin2016}%
  \BibitemOpen
  \bibfield  {author} {\bibinfo {author} {\bibfnamefont {B.~E.~K.}\
  \bibnamefont {Snodin}}, \bibinfo {author} {\bibfnamefont {F.}~\bibnamefont
  {Romano}}, \bibinfo {author} {\bibfnamefont {L.}~\bibnamefont {Rovigatti}},
  \bibinfo {author} {\bibfnamefont {T.~E.}\ \bibnamefont {Ouldridge}}, \bibinfo
  {author} {\bibfnamefont {A.~A.}\ \bibnamefont {Louis}}, \ and\ \bibinfo
  {author} {\bibfnamefont {J.~P.~K.}\ \bibnamefont {Doye}},\ }\bibfield
  {title} {\enquote {\bibinfo {title} {Direct simulation of the self-assembly
  of a small {DNA} origami},}\ }\href {\doibase 10.1021/acsnano.5b05865}
  {\bibfield  {journal} {\bibinfo  {journal} {ACS Nano}\ }\textbf {\bibinfo
  {volume} {10}},\ \bibinfo {pages} {1724} (\bibinfo {year}
  {2016})}\BibitemShut {NoStop}%
\bibitem [{\citenamefont {Snodin}\ \emph {et~al.}(2015)\citenamefont {Snodin},
  \citenamefont {Randisi}, \citenamefont {Mosayebi}, \citenamefont {Šulc},
  \citenamefont {Schreck}, \citenamefont {Romano}, \citenamefont {Ouldridge},
  \citenamefont {Tsukanov}, \citenamefont {Nir}, \citenamefont {Louis},\ and\
  \citenamefont {Doye}}]{snodin2015b}%
  \BibitemOpen
  \bibfield  {author} {\bibinfo {author} {\bibfnamefont {B.~E.~K.}\
  \bibnamefont {Snodin}}, \bibinfo {author} {\bibfnamefont {F.}~\bibnamefont
  {Randisi}}, \bibinfo {author} {\bibfnamefont {M.}~\bibnamefont {Mosayebi}},
  \bibinfo {author} {\bibfnamefont {P.}~\bibnamefont {Šulc}}, \bibinfo
  {author} {\bibfnamefont {J.~S.}\ \bibnamefont {Schreck}}, \bibinfo {author}
  {\bibfnamefont {F.}~\bibnamefont {Romano}}, \bibinfo {author} {\bibfnamefont
  {T.~E.}\ \bibnamefont {Ouldridge}}, \bibinfo {author} {\bibfnamefont
  {R.}~\bibnamefont {Tsukanov}}, \bibinfo {author} {\bibfnamefont
  {E.}~\bibnamefont {Nir}}, \bibinfo {author} {\bibfnamefont {A.~A.}\
  \bibnamefont {Louis}}, \ and\ \bibinfo {author} {\bibfnamefont {J.~P.~K.}\
  \bibnamefont {Doye}},\ }\bibfield  {title} {\enquote {\bibinfo {title}
  {Introducing improved structural properties and salt dependence into a
  coarse-grained model of {DNA}},}\ }\href {\doibase 10.1063/1.4921957}
  {\bibfield  {journal} {\bibinfo  {journal} {J. Chem. Phys.}\ }\textbf
  {\bibinfo {volume} {142}},\ \bibinfo {pages} {234901} (\bibinfo {year}
  {2015})}\BibitemShut {NoStop}%
\bibitem [{\citenamefont {Šulc}\ \emph {et~al.}(2012)\citenamefont {Šulc},
  \citenamefont {Romano}, \citenamefont {Ouldridge}, \citenamefont {Rovigatti},
  \citenamefont {Doye},\ and\ \citenamefont {Louis}}]{sulc2012}%
  \BibitemOpen
  \bibfield  {author} {\bibinfo {author} {\bibfnamefont {P.}~\bibnamefont
  {Šulc}}, \bibinfo {author} {\bibfnamefont {F.}~\bibnamefont {Romano}},
  \bibinfo {author} {\bibfnamefont {T.~E.}\ \bibnamefont {Ouldridge}}, \bibinfo
  {author} {\bibfnamefont {L.}~\bibnamefont {Rovigatti}}, \bibinfo {author}
  {\bibfnamefont {J.~P.~K.}\ \bibnamefont {Doye}}, \ and\ \bibinfo {author}
  {\bibfnamefont {A.~A.}\ \bibnamefont {Louis}},\ }\bibfield  {title} {\enquote
  {\bibinfo {title} {Sequence-dependent thermodynamics of a coarse-grained
  {DNA} model},}\ }\href {\doibase 10.1063/1.4754132} {\bibfield  {journal}
  {\bibinfo  {journal} {J. Chem. Phys.}\ }\textbf {\bibinfo {volume} {137}},\
  \bibinfo {pages} {135101} (\bibinfo {year} {2012})}\BibitemShut {NoStop}%
\bibitem [{\citenamefont {Ouldridge}, \citenamefont {Louis},\ and\
  \citenamefont {Doye}(2011)}]{ouldridge2011}%
  \BibitemOpen
  \bibfield  {author} {\bibinfo {author} {\bibfnamefont {T.~E.}\ \bibnamefont
  {Ouldridge}}, \bibinfo {author} {\bibfnamefont {A.~A.}\ \bibnamefont
  {Louis}}, \ and\ \bibinfo {author} {\bibfnamefont {J.~P.~K.}\ \bibnamefont
  {Doye}},\ }\bibfield  {title} {\enquote {\bibinfo {title} {Structural,
  mechanical, and thermodynamic properties of a coarse-grained {DNA} model},}\
  }\href {\doibase 10.1063/1.3552946} {\bibfield  {journal} {\bibinfo
  {journal} {J. Chem. Phys.}\ }\textbf {\bibinfo {volume} {134}},\ \bibinfo
  {pages} {085101} (\bibinfo {year} {2011})}\BibitemShut {NoStop}%
\bibitem [{\citenamefont {Ouldridge}, \citenamefont {Louis},\ and\
  \citenamefont {Doye}(2010)}]{ouldridge2010}%
  \BibitemOpen
  \bibfield  {author} {\bibinfo {author} {\bibfnamefont {T.~E.}\ \bibnamefont
  {Ouldridge}}, \bibinfo {author} {\bibfnamefont {A.~A.}\ \bibnamefont
  {Louis}}, \ and\ \bibinfo {author} {\bibfnamefont {J.~P.~K.}\ \bibnamefont
  {Doye}},\ }\bibfield  {title} {\enquote {\bibinfo {title} {{DNA} nanotweezers
  studied with a coarse-grained model of {DNA}},}\ }\href {\doibase
  10.1103/PhysRevLett.104.178101} {\bibfield  {journal} {\bibinfo  {journal}
  {Phys. Rev. Lett.}\ }\textbf {\bibinfo {volume} {104}},\ \bibinfo {pages}
  {178101} (\bibinfo {year} {2010})}\BibitemShut {NoStop}%
\bibitem [{\citenamefont {Reinhardt}\ and\ \citenamefont
  {Frenkel}(2014)}]{reinhardt2014}%
  \BibitemOpen
  \bibfield  {author} {\bibinfo {author} {\bibfnamefont {A.}~\bibnamefont
  {Reinhardt}}\ and\ \bibinfo {author} {\bibfnamefont {D.}~\bibnamefont
  {Frenkel}},\ }\bibfield  {title} {\enquote {\bibinfo {title} {Numerical
  evidence for nucleated self-assembly of {DNA} brick structures},}\ }\href
  {\doibase 10.1103/PhysRevLett.112.238103} {\bibfield  {journal} {\bibinfo
  {journal} {Phys. Rev. Lett.}\ }\textbf {\bibinfo {volume} {112}},\ \bibinfo
  {pages} {238103} (\bibinfo {year} {2014})}\BibitemShut {NoStop}%
\bibitem [{\citenamefont {Jacobs}, \citenamefont {Reinhardt},\ and\
  \citenamefont {Frenkel}(2015)}]{jacobs2015}%
  \BibitemOpen
  \bibfield  {author} {\bibinfo {author} {\bibfnamefont {W.~M.}\ \bibnamefont
  {Jacobs}}, \bibinfo {author} {\bibfnamefont {A.}~\bibnamefont {Reinhardt}}, \
  and\ \bibinfo {author} {\bibfnamefont {D.}~\bibnamefont {Frenkel}},\
  }\bibfield  {title} {\enquote {\bibinfo {title} {Rational design of
  self-assembly pathways for complex multicomponent structures},}\ }\href
  {\doibase 10.1073/pnas.1502210112} {\bibfield  {journal} {\bibinfo  {journal}
  {Proc. Natl. Acad. Sci. U. S. A.}\ }\textbf {\bibinfo {volume} {112}},\
  \bibinfo {pages} {6313} (\bibinfo {year} {2015})}\BibitemShut {NoStop}%
\bibitem [{\citenamefont {Reinhardt}\ and\ \citenamefont
  {Frenkel}(2016)}]{reinhardt2016}%
  \BibitemOpen
  \bibfield  {author} {\bibinfo {author} {\bibfnamefont {A.}~\bibnamefont
  {Reinhardt}}\ and\ \bibinfo {author} {\bibfnamefont {D.}~\bibnamefont
  {Frenkel}},\ }\bibfield  {title} {\enquote {\bibinfo {title} {{DNA} brick
  self-assembly with an off-lattice potential},}\ }\href {\doibase
  10.1039/C6SM01031H} {\bibfield  {journal} {\bibinfo  {journal} {Soft Matter}\
  }\textbf {\bibinfo {volume} {12}},\ \bibinfo {pages} {6253} (\bibinfo {year}
  {2016})}\BibitemShut {NoStop}%
\bibitem [{\citenamefont {Sajfutdinow}\ \emph {et~al.}(2018)\citenamefont
  {Sajfutdinow}, \citenamefont {Jacobs}, \citenamefont {Reinhardt},
  \citenamefont {Schneider},\ and\ \citenamefont {Smith}}]{sajfutdinow2018}%
  \BibitemOpen
  \bibfield  {author} {\bibinfo {author} {\bibfnamefont {M.}~\bibnamefont
  {Sajfutdinow}}, \bibinfo {author} {\bibfnamefont {W.~M.}\ \bibnamefont
  {Jacobs}}, \bibinfo {author} {\bibfnamefont {A.}~\bibnamefont {Reinhardt}},
  \bibinfo {author} {\bibfnamefont {C.}~\bibnamefont {Schneider}}, \ and\
  \bibinfo {author} {\bibfnamefont {D.~M.}\ \bibnamefont {Smith}},\ }\bibfield
  {title} {\enquote {\bibinfo {title} {Direct observation and rational design
  of nucleation behavior in addressable self-assembly},}\ }\href {\doibase
  10.1073/pnas.1806010115} {\bibfield  {journal} {\bibinfo  {journal} {Proc.
  Natl. Acad. Sci. U. S. A.}\ }\textbf {\bibinfo {volume} {115}},\ \bibinfo
  {pages} {E5877} (\bibinfo {year} {2018})}\BibitemShut {NoStop}%
\bibitem [{\citenamefont {Wayment-Steele}, \citenamefont {Frenkel},\ and\
  \citenamefont {Reinhardt}(2017)}]{wayment-steele2017}%
  \BibitemOpen
  \bibfield  {author} {\bibinfo {author} {\bibfnamefont {H.~K.}\ \bibnamefont
  {Wayment-Steele}}, \bibinfo {author} {\bibfnamefont {D.}~\bibnamefont
  {Frenkel}}, \ and\ \bibinfo {author} {\bibfnamefont {A.}~\bibnamefont
  {Reinhardt}},\ }\bibfield  {title} {\enquote {\bibinfo {title} {Investigating
  the role of boundary bricks in {DNA} brick self-assembly},}\ }\href {\doibase
  10.1039/C6SM02719A} {\bibfield  {journal} {\bibinfo  {journal} {Soft Matter}\
  }\textbf {\bibinfo {volume} {13}},\ \bibinfo {pages} {1670} (\bibinfo {year}
  {2017})}\BibitemShut {NoStop}%
\bibitem [{\citenamefont {Protozanova}, \citenamefont {Yakovchuk},\ and\
  \citenamefont {Frank-Kamenetskii}(2004)}]{protozanova2004}%
  \BibitemOpen
  \bibfield  {author} {\bibinfo {author} {\bibfnamefont {E.}~\bibnamefont
  {Protozanova}}, \bibinfo {author} {\bibfnamefont {P.}~\bibnamefont
  {Yakovchuk}}, \ and\ \bibinfo {author} {\bibfnamefont {M.~D.}\ \bibnamefont
  {Frank-Kamenetskii}},\ }\bibfield  {title} {\enquote {\bibinfo {title}
  {Stacked–unstacked equilibrium at the nick site of {DNA}},}\ }\href
  {\doibase 10.1016/j.jmb.2004.07.075} {\bibfield  {journal} {\bibinfo
  {journal} {J. Mol. Biol.}\ }\textbf {\bibinfo {volume} {342}},\ \bibinfo
  {pages} {775 } (\bibinfo {year} {2004})}\BibitemShut {NoStop}%
\bibitem [{\citenamefont {Yakovchuk}, \citenamefont {Protozanova},\ and\
  \citenamefont {Frank-Kamenetskii}(2006)}]{yakovchuk2006}%
  \BibitemOpen
  \bibfield  {author} {\bibinfo {author} {\bibfnamefont {P.}~\bibnamefont
  {Yakovchuk}}, \bibinfo {author} {\bibfnamefont {E.}~\bibnamefont
  {Protozanova}}, \ and\ \bibinfo {author} {\bibfnamefont {M.~D.}\ \bibnamefont
  {Frank-Kamenetskii}},\ }\bibfield  {title} {\enquote {\bibinfo {title}
  {Base-stacking and base-pairing contributions into thermal stability of the
  {DNA} double helix},}\ }\href {\doibase 10.1093/nar/gkj454} {\bibfield
  {journal} {\bibinfo  {journal} {Nucleic Acids Res.}\ }\textbf {\bibinfo
  {volume} {34}},\ \bibinfo {pages} {564} (\bibinfo {year} {2006})}\BibitemShut
  {NoStop}%
\bibitem [{\citenamefont {Marinari}\ and\ \citenamefont
  {Parisi}(1992)}]{marinari1992}%
  \BibitemOpen
  \bibfield  {author} {\bibinfo {author} {\bibfnamefont {E.}~\bibnamefont
  {Marinari}}\ and\ \bibinfo {author} {\bibfnamefont {G.}~\bibnamefont
  {Parisi}},\ }\bibfield  {title} {\enquote {\bibinfo {title} {Simulated
  tempering: A new {Monte} {Carlo} scheme},}\ }\href {\doibase
  10.1209/0295-5075/19/6/002} {\bibfield  {journal} {\bibinfo  {journal}
  {Europhys. Lett.}\ }\textbf {\bibinfo {volume} {19}},\ \bibinfo {pages} {451}
  (\bibinfo {year} {1992})}\BibitemShut {NoStop}%
\bibitem [{\citenamefont {Manousiouthakis}\ and\ \citenamefont
  {Deem}(1999)}]{manousiouthakis1999}%
  \BibitemOpen
  \bibfield  {author} {\bibinfo {author} {\bibfnamefont {V.~I.}\ \bibnamefont
  {Manousiouthakis}}\ and\ \bibinfo {author} {\bibfnamefont {M.~W.}\
  \bibnamefont {Deem}},\ }\bibfield  {title} {\enquote {\bibinfo {title}
  {Strict detailed balance is unnecessary in {Monte} {Carlo} simulation},}\
  }\href {\doibase 10.1063/1.477973} {\bibfield  {journal} {\bibinfo  {journal}
  {J. Chem. Phys.}\ }\textbf {\bibinfo {volume} {110}},\ \bibinfo {pages}
  {2753} (\bibinfo {year} {1999})}\BibitemShut {NoStop}%
\bibitem [{\citenamefont {Siepmann}\ and\ \citenamefont
  {Frenkel}(1992)}]{siepmann1992}%
  \BibitemOpen
  \bibfield  {author} {\bibinfo {author} {\bibfnamefont {J.~I.}\ \bibnamefont
  {Siepmann}}\ and\ \bibinfo {author} {\bibfnamefont {D.}~\bibnamefont
  {Frenkel}},\ }\bibfield  {title} {\enquote {\bibinfo {title} {Configurational
  bias {Monte} {Carlo}: A new sampling scheme for flexible chains},}\ }\href
  {\doibase 10.1080/00268979200100061} {\bibfield  {journal} {\bibinfo
  {journal} {Mol. Phys.}\ }\textbf {\bibinfo {volume} {75}},\ \bibinfo {pages}
  {59} (\bibinfo {year} {1992})}\BibitemShut {NoStop}%
\bibitem [{\citenamefont {Consta}\ \emph
  {et~al.}(1999{\natexlab{a}})\citenamefont {Consta}, \citenamefont {Wilding},
  \citenamefont {Frenkel},\ and\ \citenamefont {Alexandrowicz}}]{consta1999a}%
  \BibitemOpen
  \bibfield  {author} {\bibinfo {author} {\bibfnamefont {S.}~\bibnamefont
  {Consta}}, \bibinfo {author} {\bibfnamefont {N.~B.}\ \bibnamefont {Wilding}},
  \bibinfo {author} {\bibfnamefont {D.}~\bibnamefont {Frenkel}}, \ and\
  \bibinfo {author} {\bibfnamefont {Z.}~\bibnamefont {Alexandrowicz}},\
  }\bibfield  {title} {\enquote {\bibinfo {title} {Recoil growth: An efficient
  simulation method for multi-polymer systems},}\ }\href {\doibase
  10.1063/1.477844} {\bibfield  {journal} {\bibinfo  {journal} {J. Chem.
  Phys.}\ }\textbf {\bibinfo {volume} {110}},\ \bibinfo {pages} {3220}
  (\bibinfo {year} {1999}{\natexlab{a}})}\BibitemShut {NoStop}%
\bibitem [{\citenamefont {Consta}\ \emph
  {et~al.}(1999{\natexlab{b}})\citenamefont {Consta}, \citenamefont {Vlugt},
  \citenamefont {Hoeth}, \citenamefont {Smit},\ and\ \citenamefont
  {Frenkel}}]{consta1999b}%
  \BibitemOpen
  \bibfield  {author} {\bibinfo {author} {\bibfnamefont {S.}~\bibnamefont
  {Consta}}, \bibinfo {author} {\bibfnamefont {T.~J.~H.}\ \bibnamefont
  {Vlugt}}, \bibinfo {author} {\bibfnamefont {J.~W.}\ \bibnamefont {Hoeth}},
  \bibinfo {author} {\bibfnamefont {B.}~\bibnamefont {Smit}}, \ and\ \bibinfo
  {author} {\bibfnamefont {D.}~\bibnamefont {Frenkel}},\ }\bibfield  {title}
  {\enquote {\bibinfo {title} {Recoil growth algorithm for chain molecules with
  continuous interactions},}\ }\href {\doibase 10.1080/00268979909482926}
  {\bibfield  {journal} {\bibinfo  {journal} {Mol. Phys.}\ }\textbf {\bibinfo
  {volume} {97}},\ \bibinfo {pages} {1243} (\bibinfo {year}
  {1999}{\natexlab{b}})}\BibitemShut {NoStop}%
\bibitem [{\citenamefont {Dijkstra}, \citenamefont {Frenkel},\ and\
  \citenamefont {Hansen}(1994)}]{dijsktra1994}%
  \BibitemOpen
  \bibfield  {author} {\bibinfo {author} {\bibfnamefont {M.}~\bibnamefont
  {Dijkstra}}, \bibinfo {author} {\bibfnamefont {D.}~\bibnamefont {Frenkel}}, \
  and\ \bibinfo {author} {\bibfnamefont {J.-P.}\ \bibnamefont {Hansen}},\
  }\bibfield  {title} {\enquote {\bibinfo {title} {Phase separation in binary
  hard-core mixtures},}\ }\href {\doibase 10.1063/1.468468} {\bibfield
  {journal} {\bibinfo  {journal} {J. Chem. Phys.}\ }\textbf {\bibinfo {volume}
  {101}},\ \bibinfo {pages} {3179} (\bibinfo {year} {1994})}\BibitemShut
  {NoStop}%
\bibitem [{\citenamefont {Kim}\ \emph {et~al.}(2012)\citenamefont {Kim},
  \citenamefont {Kilchherr}, \citenamefont {Dietz},\ and\ \citenamefont
  {Bathe}}]{kim2012}%
  \BibitemOpen
  \bibfield  {author} {\bibinfo {author} {\bibfnamefont {D.-N.}\ \bibnamefont
  {Kim}}, \bibinfo {author} {\bibfnamefont {F.}~\bibnamefont {Kilchherr}},
  \bibinfo {author} {\bibfnamefont {H.}~\bibnamefont {Dietz}}, \ and\ \bibinfo
  {author} {\bibfnamefont {M.}~\bibnamefont {Bathe}},\ }\bibfield  {title}
  {\enquote {\bibinfo {title} {Quantitative prediction of 3d solution shape and
  flexibility of nucleic acid nanostructures},}\ }\href {\doibase
  10.1093/nar/gkr1173} {\bibfield  {journal} {\bibinfo  {journal} {Nucleic
  Acids Res.}\ }\textbf {\bibinfo {volume} {40}},\ \bibinfo {pages} {2862}
  (\bibinfo {year} {2012})}\BibitemShut {NoStop}%
\bibitem [{\citenamefont {Chodera}\ \emph {et~al.}(2007)\citenamefont
  {Chodera}, \citenamefont {Swope}, \citenamefont {Pitera}, \citenamefont
  {Seok},\ and\ \citenamefont {Dill}}]{chodera2007}%
  \BibitemOpen
  \bibfield  {author} {\bibinfo {author} {\bibfnamefont {J.~D.}\ \bibnamefont
  {Chodera}}, \bibinfo {author} {\bibfnamefont {W.~C.}\ \bibnamefont {Swope}},
  \bibinfo {author} {\bibfnamefont {J.~W.}\ \bibnamefont {Pitera}}, \bibinfo
  {author} {\bibfnamefont {C.}~\bibnamefont {Seok}}, \ and\ \bibinfo {author}
  {\bibfnamefont {K.~A.}\ \bibnamefont {Dill}},\ }\bibfield  {title} {\enquote
  {\bibinfo {title} {Use of the weighted histogram analysis method for the
  analysis of simulated and parallel tempering simulations},}\ }\href {\doibase
  10.1021/ct0502864} {\bibfield  {journal} {\bibinfo  {journal} {J. Chem.
  Theory Comput.}\ }\textbf {\bibinfo {volume} {3}},\ \bibinfo {pages} {26}
  (\bibinfo {year} {2007})}\BibitemShut {NoStop}%
\end{thebibliography}%
%    \newpage
    \appendix
    % appendix.tex

\section*{Appendix}
\label{sec:appendix}

\subsection{Justification for allowed kinked configurations}
\label{sec:appendix-kinks}

\begin{figure*}
    \includegraphics{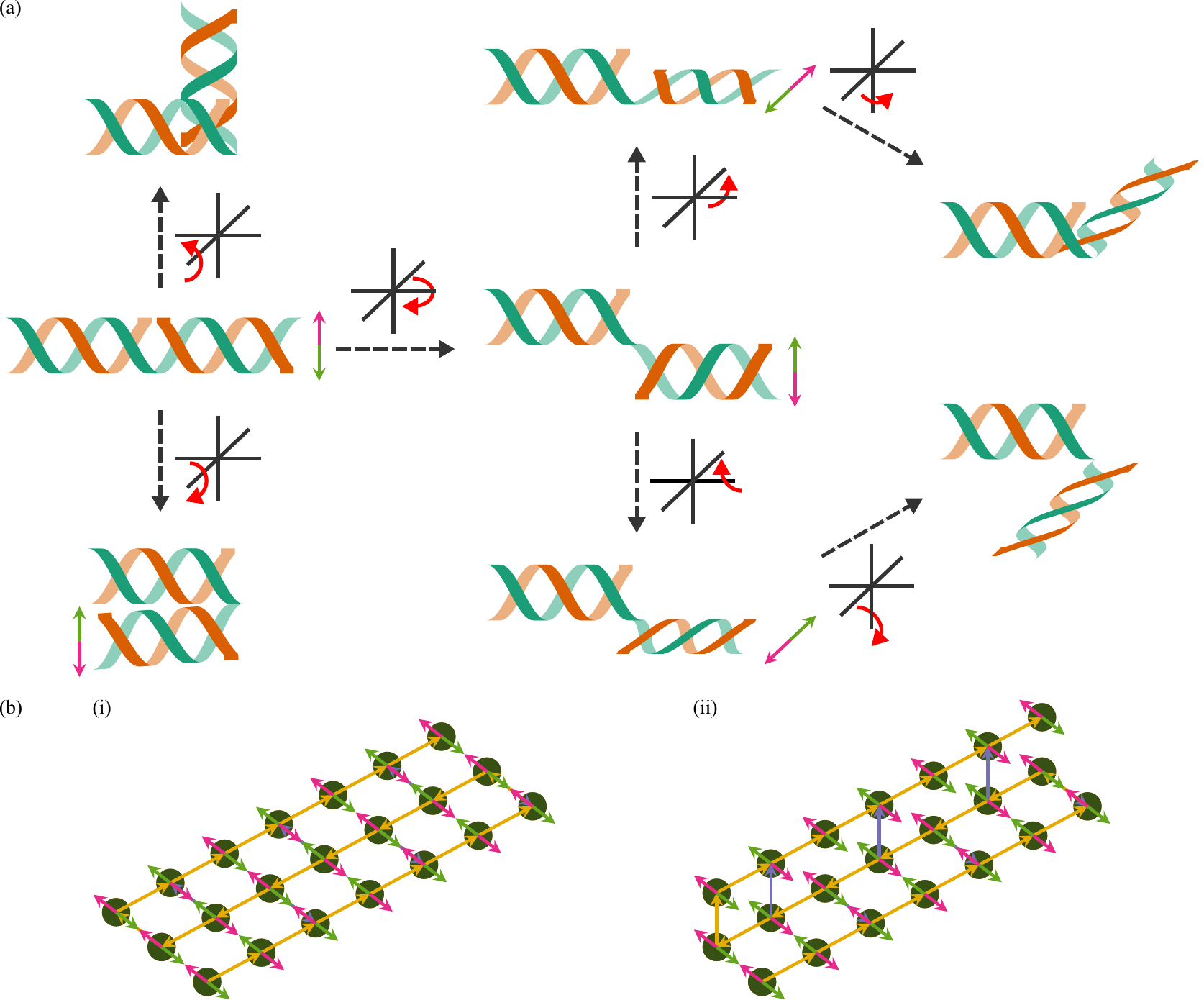}
    \caption{
        \label{fig:kinks}
        Model diagrams illustrating justification for the kink rules.
        (a) Kinks at a strand breakpoint in a helix involving 16-\ac{nt} binding domains.
        Dashed arrows show the direction of the transformation described by the associated axes and red rotation direction arrow.
        The first column shows the starting helical conformation and two conformations resulting from rotation around an axis perpendicular to both the helical axis and the first domain's orientation vector.
        The second column shows three conformations resulting from rotation around an axis parallel to the helical axis but displaced to the outside of the helix.
        The third column shows two conformations resulting from taking conformations from the second column and applying a rotation around an axis perpendicular to the helical axis and parallel to the first domain's orientation vector.
        (b) An assembled 21-binding-domain scaffold system in a planar (i) and kinked configuration (ii).
        The scaffold and staples are shown separately in \cref{fig:halftile-overview}.
    }
\end{figure*}

To understand which configurations are possible, we must consider a number of rotations of the second binding domain's helix relative to the first binding domain's helix (\cref{fig:kinks}).
To begin, consider rotating the second binding domain's helix around an axis parallel to the helical axis but displaced to the outside of the helix.
This allows for configurations in which the next-binding-domain vector of the first binding domain is perpendicular to the orientation vector of both the first and the second binding domains.
Following this first rotation with further rotations of the second binding domain's helix around an axis parallel to the orientation vector of the first will not lead to any new relative orientations of the second binding domain's orientation vector because of course the direction of the next-binding-domain vector will also be similarly rotated.
To make the model consistent with these arguments, if the first binding domain's next-binding-domain vector is perpendicular to its orientation vector, the second binding domain's orientation vector must also be perpendicular to next-binding-domain vector of the first.

\onecolumngrid
\twocolumngrid

Rotations around an axis perpendicular to the two previously mentioned rotation axes can lead to configurations in which the first binding domain's next-binding-domain vector is parallel to its orientation vector.
In the assembled state these configurations are referred to as strand crossovers.
Importantly, these configurations all have an orientation vector of the second binding domain that is equal to the orientation vector of the first binding domain.
(This also happens to be necessary if strand crossovers are to be restricted to certain intervals of base pairs)
Therefore we constrain our model to disallow kinked configurations in which the first binding domain's next-binding-domain vector is parallel to its orientation vector but not parallel to the second binding domain's orientation vector.
Configurations in which the next-binding-domain vector is antiparallel to the orientation vector of the first binding domain can also be disallowed because of steric clashes between the two binding domains.

Once we start considering kinks in which one or the other of the bound domains has its helical axis defined by virtue of the presence of an adjacent bound domain (on its chain or on one of the bound domain chains), determining which model configurations have sterically prohibited interpretations becomes even more complicated.
With the current model, it is not possible to prevent every sterically inhibited configuration without prohibitively complicated potentials.
However, only one additional rule is required to ensure that crossovers between parallel helices occur with the desired geometry.
In contrast to the rules involving only two bound domains forming a kink, the rule prohibits the underlying physical configuration by ensuring an additional kink is present, rather than directly prohibiting a model configuration.
If both of the bound domains with a kink between them have a defined helical axis, and if the two helical axes are parallel or antiparallel to each other and orthogonal to the next-binding-domain vector between the binding domains forming the kink, then there must be an additional kink present that these four bound domains are involved in and thus one less stacking interaction.

The reason why these configurations are not allowed is not clear when using the idealized cartoon helix representation.
It is best explained by considering parallel helices with multiple crossovers.
Consider a 21-binding-domain scaffold system that is designed to assemble into three parallel helices with crossovers between them (\cref{fig:halftile-overview}).
Without this rule, \cref{fig:kinks}(b)(i) and \cref{fig:kinks}(b)(ii) would be considered energetically equivalent, yet we know that the geometry of \cref{fig:kinks}(b)(ii) is unfavourable.
If it were not, the classic tile structures (this 21-binding-domain system is a subset of the rectangular tile of ref.~\citenum{rothemund2006}) would lack rigidity in solution, which is not the case~\cite{kim2012}.

\subsection{Numerical validation of move types}
\label{sec:appendix-validation}

\begin{figure}
    \centering
    \includegraphics{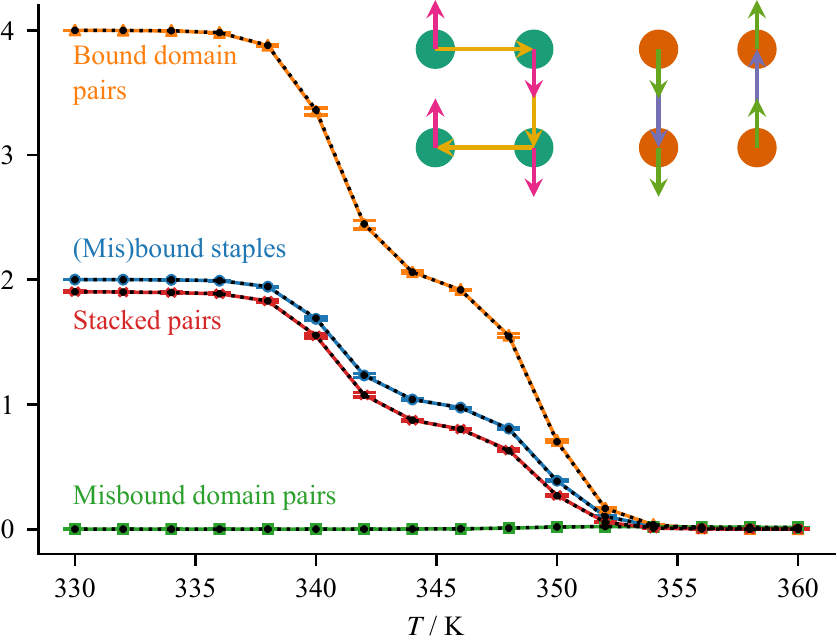}
    \caption{
        \label{fig:validation}
        Numerical validation of the move types and their implementation.
        Plotted are the mean order parameters against temperature.
        The exact result for each order parameter is plotted in dashed black lines.
        The error bars on the simulation results represent the standard error in the means across ten independent simulations.
        Simulations were run with the same parameters as the simulations referred to in \cref{fig:snodin-melting}.
        Above the curves is a schematic representation of the four-binding-domain scaffold system used.
        The scaffold and staples are shown in assembled configurations, but for clarity have been drawn separately.
    }
\end{figure}

To examine the validity of the move types, simulations of a toy system were run and compared to exact results.
The system used consists of a four-binding-domain scaffold and two staples, in which one of the staples links the terminal scaffold binding domains in the assembled state, as can be seen in \cref{fig:validation}(a)).
The exact results were calculated by taking the ensemble averages across all configurations that have at most four total staples or two staples of a given type, which were determined with a recursive enumeration algorithm.
The move set was nearly the same as that used to run the simulations presented in the results section with the exception that the maximum number of scaffold domains to regrow is set to four.
The average number of bound domain pairs, the average number of (mis)bound staples, the average number of misbound domain pairs, and the average number of stacked binding domain pairs for both the simulation and enumeration results are plotted in \cref{fig:validation}, which clearly shows that the two approaches agree within sampling error.

\subsection{Optimization of move sets}
\label{sec:appendix-optimization}

A given move set has many free parameters and cannot be fully optimized without extensive efforts.
For the move set used in the current study, which has an orientation rotation move type, a staple exchange move type, a \ac{CB} staple regrowth move type, a contiguous \ac{CT}\ac{RG} scaffold regrowth move type, a non-contiguous \ac{CT}\ac{RG} scaffold regrowth move type, and a \ac{REMC} exchange move type, there are 12 adjustable parameters.
Further, the optimal parameters will be different depending on the system being simulated, as well as the simulation conditions and model parameters.
We instead undertake only a small amount of optimization to avoid wasting effort on the diminishing returns associated with more thorough optimization.
The strategy was to optimize parameters in isolation by assuming that the dependency in optimal value of a given parameter on the others is small.

The individual parameters were optimized by running simulations on the 24-binding-domain scaffold system, from which we determined the mean times to the first assembled state and to the first fully stacked assembled state, as well as estimated the effective sample size~\cite{chodera2007}.
Based on this, for both the contiguous and non-contiguous \ac{CT}\ac{RG} scaffold regrowth move types, we chose a maximum of one recoil, a maximum of 36 (all possible) configurations to be attempted at each growth step, and a maximum of 12 total scaffold binding domains to attempt to regrow.
For the non-contiguous \ac{CT}\ac{RG} move type, we chose a maximum of two scaffold binding domains to attempt to regrow per segment.
For the ratio of move type frequencies, we chose orientation rotation, staple exchange, staple regrowth, contiguous \ac{CT}\ac{RG} scaffold regrowth, and non-contiguous \ac{CT}\ac{RG} scaffold regrowth moves in a ratio of $2\mathbin{:}1\mathbin{:}1\mathbin{:}1\mathbin{:}1$.
Finally, we make an exchange attempt between replicas every 100 steps.
For many of the parameters, there was a substantial range in which the sampling efficiency was very similar, so many of the values given above could well have been chosen differently.

    \clearpage
    % supinfo.tex

\renewcommand{\thefigure}{S\arabic{figure}}
\renewcommand*{\theHfigure}{\thepart.\thefigure}  % this is to prevent hyperref labels from pointing to the wrong figure (since we start numbering them from 0 again)
\setcounter{figure}{0}
\pagenumbering{S\arabic}
\renewcommand*{\thepage}{S\arabic{page}}

\onecolumngrid
\section*{Supplementary figures}
\twocolumngrid

\begin{figure}[h]
    \centering
    \includegraphics{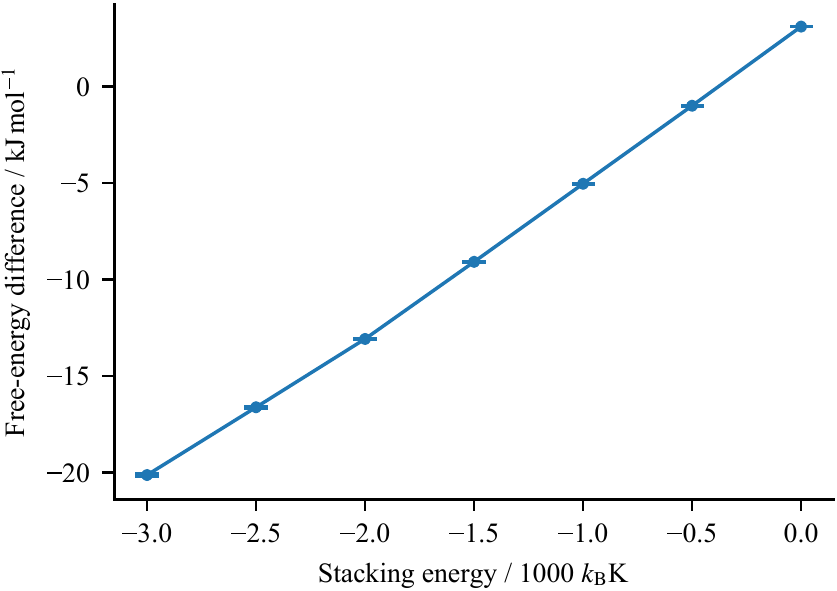}
    \caption{
        \label{fig:stacking-enthalpy}
        Stacking free energy as a function of stacking energy calculated from simulations of a two-binding-domain scaffold system with two one-binding-domain staples.
        The simulations were run at \SI{320}{\kelvin}, which is below the melting temperature.
        The error bars represent the standard error in the means across three independent simulations.
        Simulations were run with a staple concentration of \SI{100}{\nano\Molar} and a monovalent cation concentration of \SI{0.5}{\Molar}.
    }
\end{figure}

\newpage

\begin{figure}[h]
    \centering
    \includegraphics{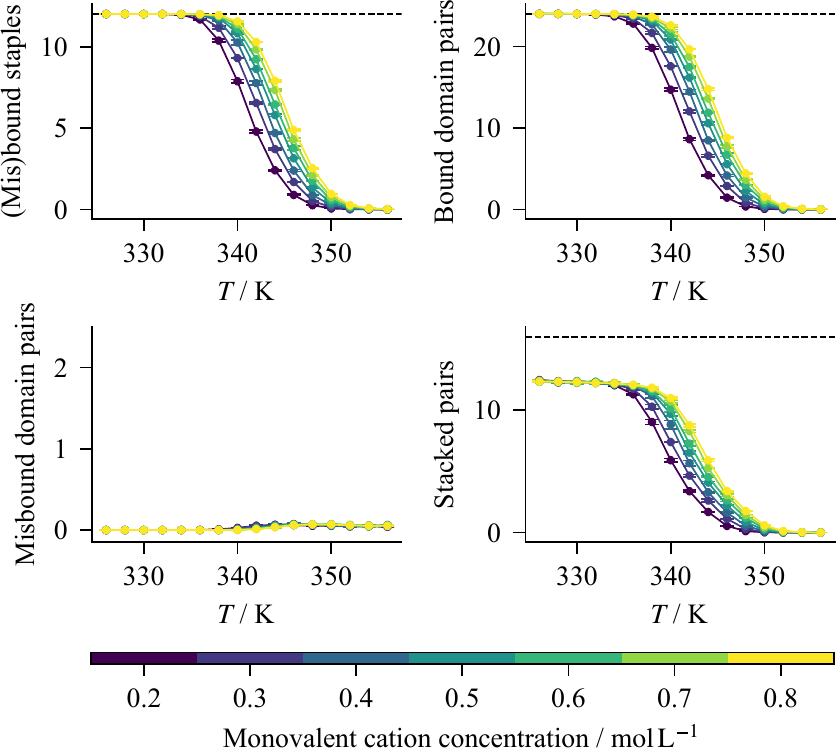}
    \caption{
        \label{fig:cation-melt}
        Mean order parameters from simulations of the 24-binding-domain system plotted against temperature for a range of sodium ion concentrations.
        The dashed lines correspond to the expected order parameter values in the fully assembled or fully stacked configurations.
        The error bars represent the standard error in the means across three independent simulations.
        Simulations were run with a staple concentration of \SI{100}{\nano\Molar} and a stacking energy of $-1000\,k_\text{B}\si{\kelvin}$.
    }
\end{figure}

\onecolumngrid

\begin{figure}[h]
    \includegraphics{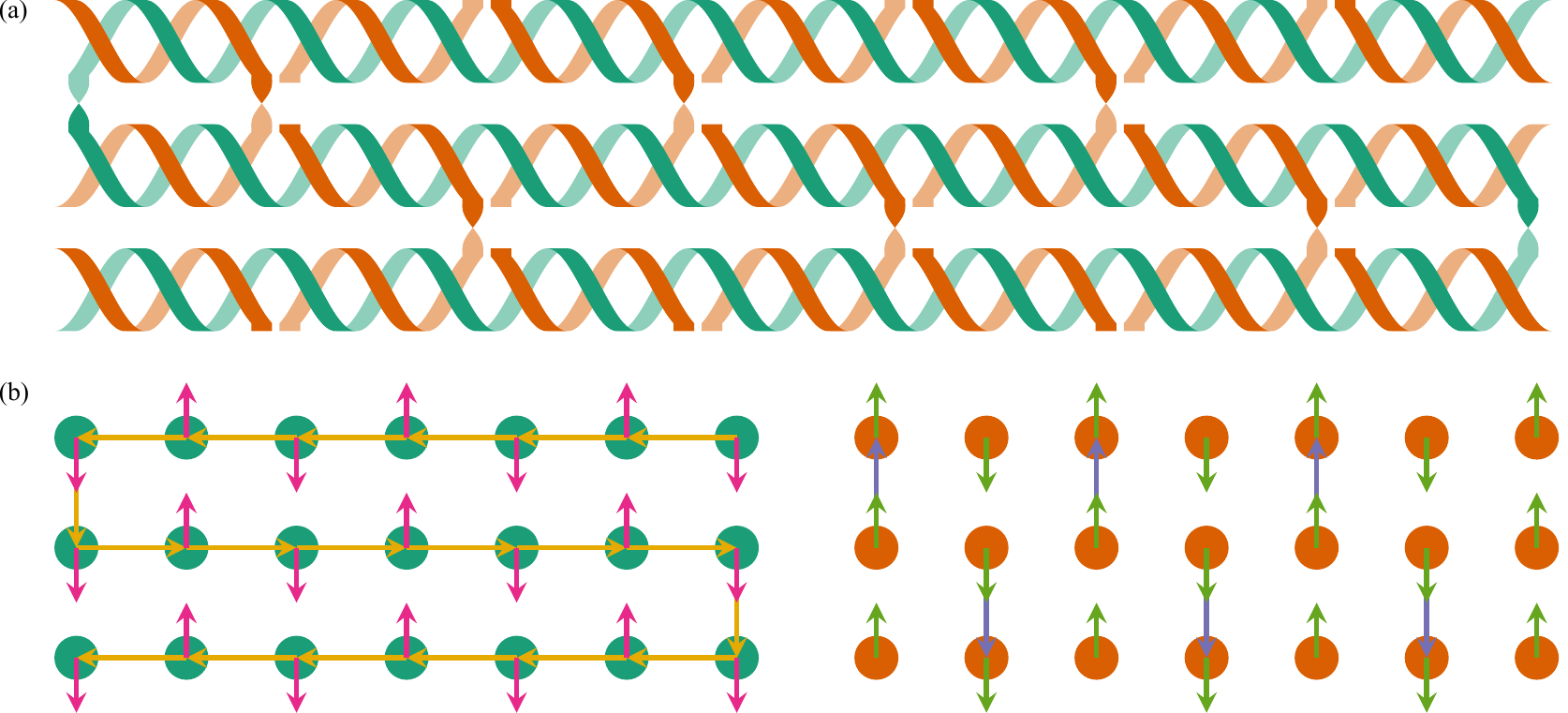}
    \caption{
        \label{fig:halftile-overview}
        Schematic representations of the 21-binding-domain system.
        (a) Helical cartoon representation of the system in a fully assembled, planar configuration.
        (b) Representation of the system with the proposed model.
        The scaffold and staples are shown in assembled configurations, but for clarity have been drawn separately.
    }
\end{figure}

\end{document}